\documentclass[sn-mathphys-num]{sn-jnl}

\usepackage{amsmath}
\usepackage{amsfonts}
\usepackage{algorithmic}
\usepackage{graphicx}
\usepackage{textcomp}
\usepackage{xcolor}
\usepackage{multirow}
\usepackage{multicol}
\usepackage{pgfplots}
\usepackage{array}
\usepackage{pgfplots}
\usepackage{caption}
\usepackage{subcaption}
\usepackage{booktabs}
\usepackage{makecell}
\usepackage{framed}
\usepackage{listings}
\usepackage{array}
\usepackage{lineno}
\usepackage{fixltx2e}
\usepackage{graphicx}
\usepackage{framed}
\usepackage{enumitem}
\usepackage{pgf-pie}
\usetikzlibrary{patterns}
\usetikzlibrary{pgfplots.statistics}
\usepackage{csquotes}
\usepackage{relsize}
\usepackage{float}
\usepackage{tabularx}
\usepackage{algorithmic}
\usepackage{graphicx}
\usepackage{textcomp}
\usepackage{xcolor}
\usepackage{hyperref}
\usetikzlibrary{arrows}
\usepackage{colortbl}
\usetikzlibrary{positioning, arrows.meta}
\usepackage[skins]{tcolorbox}
\usepackage{venndiagram}
\usepackage{adjustbox}
\usepackage{MnSymbol,bbding,pifont}
\usepackage{float}
\usepackage{longtable}

\definecolor{info}{RGB}{228, 155, 15} 
\definecolor{negative}{RGB}{220, 20, 60} 
\definecolor{positive}{RGB}{34, 139, 34} 
\definecolor{light}{RGB}{243,246,244}

\definecolor{calmblue}{HTML}{39A7E7}
\definecolor{lightblue}{HTML}{e7f3fe}

\definecolor{lightgreen}{HTML}{00C805}
\definecolor{lightergreen}{HTML}{ecfbe7}
\definecolor{dgreen}{HTML}{38761d}
\definecolor{darkgreen}{HTML}{004022} 

\definecolor{yellow}{HTML}{FFC629}
\definecolor{lightyellow}{HTML}{FFF4DA}
\definecolor{darkyellow}{HTML}{F9B205}

\definecolor{pinkish}{HTML}{FF00BF} 
\definecolor{lightpink}{HTML}{ffeff4}

\definecolor{myOrange}{RGB}{245, 125, 50}
\definecolor{grey6}{HTML}{EBEEF0}

\definecolor{tinderPink}{HTML}{FD267A} 
\definecolor{tinderOrange}{HTML}{FF7854} 
\colorlet{tindercolor}{tinderPink!65!tinderOrange}

\def\BibTeX{{\rm B\kern-.05em{\sc i\kern-.025em b}\kern-.08em
		T\kern-.1667em\lower.7ex\hbox{E}\kern-.125emX}}

\def\EntityDensityApps{Uber/0.0451/0.0833/true/black/grey6/black, 
	Lyft/0.0344/0.065/false/pinkish/lightpink/pinkish, 
	Tinder/0.0714/0.083/false/tindercolor/lightpink/tindercolor, 
	Bumble/0.0806/0.106/false/darkyellow/lightyellow/darkyellow, 
	Robinhood/0.034/0.076/true/dgreen/lightergreen/dgreen, 
	Acorn/0.058/0.091/false/lightgreen/lightergreen/lightgreen, 
	Calm/0.032/0.050/false/calmblue/lightblue/calmblue, 
	Headspace/0.025/0.067/false/myOrange/lightyellow/myOrange}

\newcommand{\densityplot}[9]{
	\begin{subfigure}{0.225\textwidth}
		\begin{tikzpicture}
			\begin{axis}[
				xlabel={},
				ylabel={},
				label style={
					font=\footnotesize
				},
				ticklabel style={
					font=\footnotesize
				},
				grid=none, 
				width=1.38\textwidth, 
				height=120pt,
				ymin=0,
				ymax=0.2,
				ytick distance=0.03,
				xtick={2, 3, 4, 5}, 
				xmin=1, 
				xmax = 5,
				extra x ticks={1}, 
				extra x tick labels={1}, 
				ylabel style = {
					yshift = -25pt,
					xshift = 2pt,
				},
				tick style={
					draw=none,
				},
				yticklabel={
					\ifthenelse{\boolean{#6}}
					{\pgfmathprintnumber{\tick}}
					{}
				},
				yticklabel style={
					/pgf/number format/fixed,
					/pgf/number format/precision=2,        
				},
				title={\textcolor{#7}{{#2}}}, 
				title style={
					at={(0.02,0.89)}, 
					anchor=north west,
					draw={#9}, 
					fill={#8}, 
					font=\tiny,
					align=left,
					inner sep=1.8pt
				}
				]
				
				\addplot[black,mark=*, mark options={scale=0.5}] table[x=X, y=Y, col sep=comma] {#1}; 
				\addplot[black,mark=square, mark options={scale=0.5}] table[x=X, y=Y, col sep=comma] {#3}; 
				\addplot[black, mark=none, dashed, const plot] coordinates {#4}; 
				\addplot[black, mark=none, dotted, thick, const plot] coordinates {#5}; 
				
			\end{axis}
			
		\end{tikzpicture}
	\end{subfigure}
}

\begin{document}

\title[Article Title]{Mobile Application Review Summarization using Chain of Density Prompting}

\author*[1]{\fnm{Shristi} \sur{Shrestha}}\email{sshre35@lsu.edu}

\author*[2]{\fnm{Anas} \sur{Mahmoud}}\email{mahmoud@csc.lsu.edu}
\equalcont{These authors contributed equally to this work.}

\affil*[1,2]{\orgdiv{Division of Computer Science and Engineering}, \orgname{Louisiana State University}, \orgaddress{\city{Baton Rouge}, \state{LA}, \country{USA}}}


\abstract{Mobile app users commonly rely on app store ratings and reviews to find apps that suit their needs. However, the sheer volume of reviews available on app stores can lead to information overload, thus impeding users' ability to make informed app selection decisions. To address this challenge, we leverage Large Language Models (LLMs) to summarize mobile app reviews. In particular, we use the Chain of Density (CoD) prompt to guide OpenAI GPT-4 to generate abstractive, semantically dense, and easily interpretable summaries of mobile app reviews. The CoD prompt is engineered to iteratively extract salient entities from the source text and fuse them into a fixed-length summary. We evaluate the performance of our approach using a large dataset of mobile app reviews. We further conduct an empirical evaluation with 48 study participants to assess the readability of the generated summaries. Our results demonstrate that adapting the CoD prompt to focus on app features improves its ability to extract key themes from user reviews and generate natural language summaries tailored for end-user consumption. The prompt also manages to maintain the readability of the generated summaries while increasing their semantic density. Our work in this paper aims to improve mobile app users' experience by providing an effective mechanism for summarizing important user feedback in the review stream.}




\maketitle
\section{Introduction}
The explosive growth of mobile application (app) stores over the past decade has left app users with virtually unlimited options to choose from. With thousands of apps classified under each app store category, locating high-quality apps that meet specific user needs can be a challenging task. To help app users make informed app selection decisions, app stores support internal mechanisms for app rating. Apps that receive more reviews and have higher average star ratings are perceived to be higher in quality and, thus, are more likely to be installed~\cite{Siegfried15,Dogruel15}.

A main limitation of existing app review systems is that reviews are displayed in a list format, often ranked by date or helpfulness votes (\textit{likes}). However, popular apps frequently receive hundreds of new reviews per day, adding up to thousands of reviews per application domain~\cite{Jha18}. At such a scale, app users may experience information overload, a phenomenon where decision-makers face more information than they can process, thus hindering their ability to make optimal decisions~\cite{Ackoff67,Bawden09}. 

\begin{figure} [ht]
	\centering
	\includegraphics[trim={0 0cm 0 1cm}, scale=0.17]{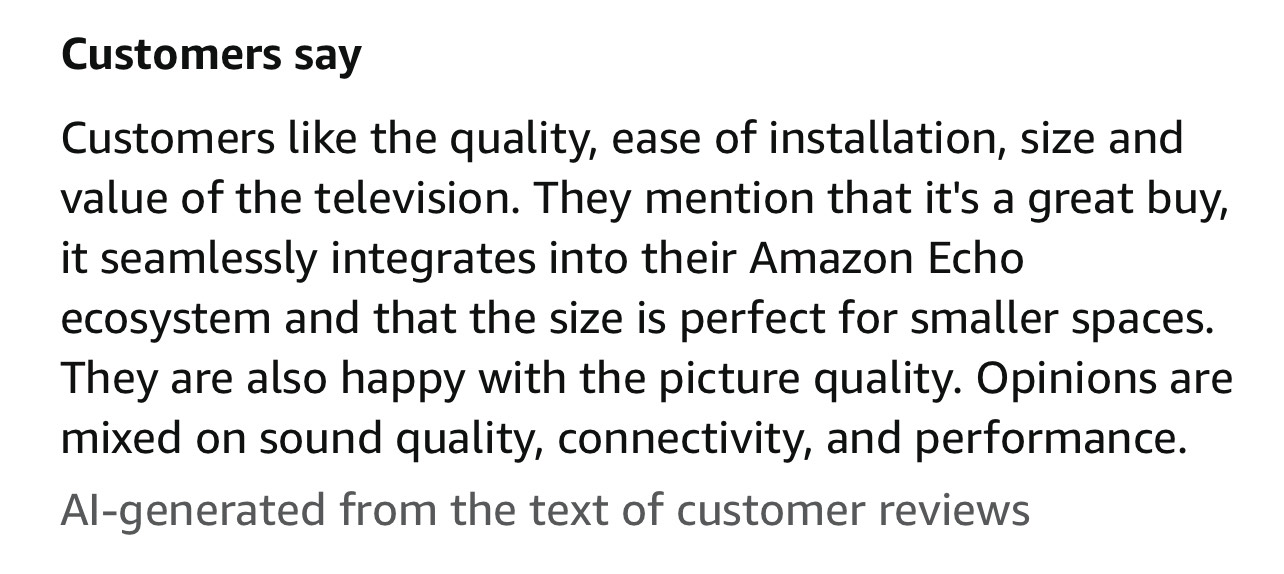}
	\centering
	\caption{An AI-generated review summary of a smart TV listing on Amazon.}
	\label{Fig:AmazonS}
\end{figure}

The problem of review overload has long affected online retail platforms~\cite{Park08,Furner16}. Best selling products on sites such as Amazon and eBay receive thousands of reviews per day. The sheer volume of this information has been linked to consumer confusion and indecision~\cite{Lappas12,Hu19}. To address this challenge, AI-generated summaries of product reviews are increasingly used. These summaries (e.g., Fig.~\ref{Fig:AmazonS}) aim to help consumers quickly grasp the key points across many reviews, distilling the essence of other buyers' opinions without requiring them to read individual reviews.

Similar to online retail, summarization has been proposed as a strategy to address the large volume of app reviews available on mobile app stores~\cite{Jha18, Ebrahimi23}. However, the proposed techniques commonly target app developers rather than their end-users. Thus, generated summaries are often extractive, unstructured, and mainly geared towards technical issues, such as bug reports and feature requests~\cite{Tao20,Chen14,Sorbo16,Jha18}. To address these limitations, this paper aims to generate abstractive app review summaries that are semantically dense, yet easily comprehensible to the average app user. The objective is to enable prospective users to identify salient themes in the review stream and effectively locate mobile apps that meet their specific needs.

To achieve our goal, we leverage Large Language Models (LLMs). Over the past few years, LLMs have had significant impact on the field of automatic text summarization~\cite{Adams23,Liu23,Shin23}. Unlike classical rule-based or frequency-based summarization techniques, contemporary language models such as OpenAI GPT-4 can be prompted to generate high-quality human-like summaries of text without the need for large training datasets or exhaustive parameter tuning~\cite{Llewellyn14,Khabiri11}. Building upon these capabilities, in this paper, we employ the Chain of Density (CoD) prompt to summarize mobile app reviews~\cite{Adams23}. This prompt is engineered to guide LLMs to iteratively enhance the semantic density of generated summaries while maintaining control over their naturalness and length. To evaluate the effectiveness of the CoD prompt within the realm of mobile app reviews, we compare its performance against extractive summarization and vanilla prompting as experimental baselines. Additionally, we conduct a user study to assess the readability of generated summaries. Our objective is to investigate the capabilities of the CoD prompt as a practical solution for generating natural and semantically dense summaries of mobile app reviews that are specifically intended for end-user consumption. In particular, in this paper, we advance the state of research and practice as follows: 

\begin{itemize}
\item We propose and empirically evaluate an adapted version of the Chain of Density (CoD) prompt, specifically designed to handle the unique structure and format of mobile app reviews. Our results demonstrate that the iterative nature of the CoD prompt produces summaries with greater semantic density compared to extractive methods or summaries generated using a vanilla prompt. Additionally, we investigate and quantify how the density and structure of generated summaries vary across different LLMs.

\item We conduct a user study with 48 subjects to assess the readability of generated summaries at the different iterations of the CoD prompt. Our results show that the CoD prompt can maintain the readability and naturalness of generated summaries even at higher iterations of the summarization process.  

\item We propose a strategy for integrating mobile app review summaries into existing app store rating systems and we discuss how such summaries can help mitigate common consumer biases, such as anchoring and negativity bias.
\end{itemize}

The remainder of this paper is organized as follows: In Section~\ref{sec:bg}, we review existing work on app review summarization and identify its main limitations. Section~\ref{sec:approach} presents the CoD prompt, formulates our research questions, and describes our dataset. Section~\ref{sec:eval} details our evaluation procedure and presents the results. Section~\ref{sec:discuss} discusses the implications of our findings and outlines integration and enhancement strategies. In Section~\ref{sec:threats}, we address threats to the validity of our study. Finally, Section~\ref{sec:conc} concludes the paper and discusses directions for future work.

\section{Existing Work and Limitations}
\label{sec:bg}
Text summarization aims to identify the dominant themes in a source text and represent them cohesively and concisely, effectively generating a meaningful summary~\cite{Llewellyn14,Khabiri11}. Summarization techniques can be either extractive or abstractive. Extractive techniques select important themes from the source text as verbatim excerpts and stitch them together to construct a summary. Themes to be included in the summary are typically identified using statistical models of word-frequency. Abstractive summarization techniques, on the other hand, aim to generate natural descriptions of the main themes in a source text. Therefore, lexical parsing and paraphrasing are required to form new sentences that convey the essential information in the source text~\cite{Hahn00,Jin24}. 

Summarization has been commonly used in the context of mobile app reviews. The objective is to bring app developers' attention to the pressing issues that need to be addressed in the next release of the app~\cite{Jha18}. For instance, Jha and Mahmoud~\cite{Jha18} employed several extractive summarization techniques to summarize informative mobile app reviews. The results showed that SumBasic~\cite{Vanderwende07}, a frequency-based approach with redundancy control, effectively captured most concerns in app user reviews, achieving an average recall of 71\%. Hybrid TF.IDF was also competitive, with an average recall of 60\%. However, LexRank~\cite{Erkan04} \textemdash a graph-based summarization algorithm \textemdash failed to compete, recalling only 41\% of users' concerns from the reviews. 

Chen et al.~\cite{Chen14} presented AR-Miner, a computational framework for app review mining. AR-miner used a semi-supervised classifier to categorize reviews into informative and non-informative feedback. Informative reviews were then grouped and prioritized using topic modeling. Extensive experiments and case studies demonstrated the effectiveness of AR-Miner in summarizing user reviews with minimal manual effort. 

Sorbo et al.~\cite{Sorbo16} introduced SURF, an approach to summarize maintenance information in app reviews. SURF relied on a conceptual model of user intention and review topic classification to identify sentences in reviews that contain important technical feedback. An end-to-end evaluation over a dataset of user reviews showed that SURF was able to achieve high levels of accuracy as well as help app developers plan for future software changes.  

Tao et al.~\cite{Tao20} introduced SRR-Miner, a summarization approach that automatically extracted security-related sentences from mobile app reviews. SRR-Miner leveraged predefined semantic patterns to extract users' sentiments using misbehavior, aspect, and opinion words in review sentences. Evaluating SRR-Miner on a dataset of 17 apps showed that it outperformed several existing machine learning baselines, especially in summarizing security and privacy reviews. 

More recently, Ebrahimi et al.~\cite{Ebrahimi23} proposed an unsupervised approach for summarizing privacy concerns in mobile app reviews. The proposed approach initially constructed catalogs of privacy seed words in various application domains. These seeds were then used to guide extractive summarization algorithms toward privacy issues in user reviews. The proposed approach was evaluated on a dataset of 2.6 million app reviews. The results showed that seeding summarization algorithms helped them capture privacy concerns in app reviews with minimum redundancy. 

Despite these advances, existing app review summarization techniques are limited in terms of practicality and structure. For instance, most of the proposed techniques are extractive, resulting in review summaries that often appear unnatural and lack structure~\cite{Tao20,Chen14,Jha18}. This is because these summaries are primarily intended for app developers, aiming to help them identify critical concerns in their app reviews with less emphasis on readability and structure. Furthermore, existing summarization techniques rely on word frequency analysis to identify important themes in user reviews and control for redundancy. Thus, sparse themes tend to be either missed or underrepresented in generated summaries. Catalog-assisted summarization techniques (e.g.,~\cite{Ebrahimi23,Tao20}) can alleviate this problem by seeding summarization algorithms with words that point them to rare themes in user reviews. However, constructing and maintaining domain-specific catalogs for each specific user concern in each mobile app category can be a tedious process. Another limitation stems from the fact that user reviews are short and often expressed using informal, syntactically incorrect, and semantically restricted jargon. Extractive summarization algorithms, such as Hybrid TF.IDF and SumBasic, and topic modeling techniques, such as LDA, are susceptible to incorporating such informal text in the final summary, thus generating poorly readable or semantically incoherent review summaries.   

To address these limitations, in this paper, we leverage Large Language Models (LLMs) to generate abstract and natural summaries of mobile app reviews~\cite{Goyal23}. In particular, we propose to use the Chain of Density (CoD) prompt to craft contextually relevant, semantically dense, yet easily comprehensible summaries of app reviews~\cite{Adams23}. Our assumption is that the iterative nature of the CoD prompt can help overcome the inherent semantic limitations of app reviews while maintaining the length and readability of generated summaries. In what follows, we describe our approach, assumptions, and research questions in detail.
 
\section{Approach}
\label{sec:approach}
In this section, we present the Chain of Density (CoD) prompt, formulate our research questions, and describe our experimental dataset.  

\subsection{The Chain of Density Prompt}
Proposed by Adams et al.~\cite{Adams23}, the CoD prompt directs LLMs to generate an initial sparse summary of dominant entities in news articles. An entity refers to a specific piece of information or concept that can be recognized in the text. For example, an entity in a news article could refer to peoples' names, locations, organizations, dates, or events. The prompt then iteratively fuses more entities from the source text into the summary but without increasing its length. The objective is to produce abstractive summaries that exhibit more fusion and are less biased towards dominant themes. 

The CoD prompt is shown in Fig. 2. A missing entity is defined as \textit{``relevant''} to the main story, \textit{``specific''} yet concise, \textit{``novel''} or not included in the previous summary, \textit{``faithful''} or actually exists in the source text, and \textit{``anywhere''} in the text. Semantic density is imposed using the instructions \textit{``removal of uninformative phrases''} and \textit{``concise yet self-contained''}, while abstractiveness is enforced through the instructions \textit{``improve flow''} and \textit{``easily understood.''} Length is controlled through the \textit{``80 words''} constraint. This restriction is necessary to prevent generating excessively long summaries~\cite{Liu23}.

\begin{figure}[ht]
	\begin{tcolorbox}[colback=gray!5,colframe=gray!40!black,]
		\scriptsize
			
		\texttt{Article: ARTICLE}\\
		
		\texttt{You will generate increasingly concise, entity-dense summaries of the above article.}\\
		
		\texttt{\textbf{Repeat the following 2 steps 5 times.}}
		
		\begin{enumerate}
			\item \texttt{Identify 1-3 informative entities (";" delimited) from the article which are missing from the previously generated summary.}
			\item \texttt{Write a new, denser summary of identical length which covers every entity and detail from the previous summary plus the missing entities.}
		\end{enumerate}
		
		\textbf{\texttt{A missing entity is:}}
		
		\begin{itemize}
			\item \texttt{relevant to the main story,}
			\item \texttt{specific yet concise (5 words or fewer),}
			\item \texttt{novel (not in the previous summary),}
			\item \texttt{faithful (present in the article),}
			\item \texttt{anywhere (can be located anywhere in the article).}
		\end{itemize}
		
		\textbf{\texttt{Guidelines:}}
		
		\texttt{The first summary should be long (4-5 sentences, ~80 words) yet highly non-specific, containing little information beyond the entities marked as missing. Use overly verbose language and fillers (e.g., "this article discusses") to reach ~80 words. Make every word count: rewrite the previous summary to improve flow and make space for additional entities. Make space with fusion, compression, and removal of uninformative phrases like, "the article discusses." The summaries should become highly dense and concise yet self-contained, i.e., easily understood without the article. Missing entities can appear anywhere in the new summary. Never drop entities from the previous summary. If space cannot be made, add fewer new entities. Remember, use the same number of words for each summary." }
	\end{tcolorbox}
	\caption{The CoD prompt as proposed by Adam et al.~\cite{Adams23}}
	\label{fig:cod}
\end{figure}

\subsection{Research Questions}
In this paper, we utilize the Chain of Density (CoD) prompt to solicit increasingly dense, informative, and easily interpretable summaries of mobile app reviews. The underlying assumption is that the iterative nature of the CoD prompt can force the LLM to recall more salient entities from the reviews, thus overcome the limitations of frequency-based text summarization and modeling techniques~\cite{Jha18,Jin24,Sorbo16}. To guide our analysis, we formulate the following research questions:

\begin{itemize}
	\item \textit{\textbf{RQ$_1$: Can the CoD prompt solicit meaningful summaries of mobile app reviews?}}\\
In this research question, we explore whether the CoD prompt is effective for summarizing short, informal, and unstructured review texts. Mobile app users often express their concerns using varied language~\cite{Ebrahimi23, Jha18}. While the CoD prompt has demonstrated strong performance in summarizing structured text, its ability to handle ambiguous language and polysemy in user reviews remains unclear. To address this, we compare CoD generated summaries with those produced by a simpler frequency-based summarization technique and vanilla prompting. If a vanilla prompt can generate equally dense and readable summaries, adopting a relatively complex prompt like the CoD may be unnecessary.
	
	\item \textit{\textbf{RQ$_2$: Can the CoD summaries maintain readability as their semantic density increases?}}\
An effective abstractive summarization method should generate summaries that are both informative and easily readable. To evaluate the readability of CoD generated summaries, under this research question, we conduct a human study to explore whether the summaries remain interpretable as their semantic density increases.

	\item \textit{\textbf{RQ$_3$: How does the performance of the CoD prompt vary across different LLMs?}}\
This research question aims to investigate how the density and readability of CoD generated review summaries differ across various LLMs. Our goal is to quantify these performance variations and provide insights to guide prompt design for diverse applications and contexts, ultimately improving cross-model generalizability.
\end{itemize}

\subsection{Data}
To answer our research questions, we select eight popular apps from four application domains (two apps from each domain) to include in our dataset. We define a domain as a set of apps that support relatively similar core functionality. To conduct our analysis, we select the domains of ride-hailing, online dating, investing, and  mental health. Apps in these domains vary substantially in their functionality and target-user populations, thus can enhance the external validity of our findings. The apps from each domain were selected based on their App Store rankings by the time of conducting the study. Apple compiles several metrics (installs, active users, reviews, etc.) to determine app rankings in different categories. We considered the two highest ranked apps in each category. These apps have massive userbases, receive thousands of reviews weekly, and users often find themselves choosing between them (e.g., Uber vs. Lyft). 

Table~\ref{Tab:apps_popularity} shows the overall ratings, ranks, and number of user reviews for the apps included in our dataset based on the information available on the Apple App Store in December 2023. We collected 37,245 reviews posted after 2020 using the AppStore scraper library. Including only recent reviews ensures our analysis produces timely summaries. The dataset, along with our replication scripts, are available in the supplemental material.

\begin{table*}[ht]
	\centering
	\footnotesize
	\renewcommand{\arraystretch}{1.2}
	\caption{A description of the sample apps included in our analysis, showing their ratings, ranks in their categories, total number of reviews, and the number of reviews posted after 2020.}
	\smallskip 
	\begin{tabular}{p{0.12\textwidth} p{0.1\textwidth} c l c c}
		\Xhline{1.5\arrayrulewidth}
		\Xhline{1.5\arrayrulewidth}
		\textbf{Domain} & \textbf{App} & \textbf{Rating} & \textbf{Rank} & \textbf{Total Reviews} & \textbf{Reviews ($>$ 2020)} \\
		\Xhline{1.5\arrayrulewidth}
		\Xhline{1.5\arrayrulewidth}
		\multirow{2}{*}{\rotatebox{0}{\parbox{1.5cm}{Ride-hailing}}} & Uber & 4.9 & \#1 Travel & 7.8M & 5,018 \\
		& Lyft & 4.9  & \#3 Travel & 13.6M & 6,789 \\
		\hline
		\multirow{2}{*}{\rotatebox{0}{\parbox{1.8cm}{Online Dating}}} & Tinder & 4.0 & \#8 Lifestyle & 840K & 5,004  \\
		& Bumble & 4.3 & \#12 Lifestyle & 1.4M & 3,245 \\
		\hline
		\multirow{2}{*}{\rotatebox{0}{\parbox{1cm}{Investing}}} & Robinhood & 4.2 & \#27 Finance & 4.2M & 5,009 \\
		& Acorn & 4.7 & \#44 Finance & 880K & 2,813  \\
		\hline
		\multirow{2}{*}{\rotatebox{0}{\parbox{1.8cm}{Mental Health}}} & Calm & 4.8 & \#16 Health \& Fitness & 1.7M  & 4,908  \\
		& Headspace & 4.8 & \#55 Health \& Fitness & 940K & 4,459  \\
		\Xhline{1.5\arrayrulewidth}

		
		\Xhline{1.5\arrayrulewidth}
	\end{tabular}
	\label{Tab:apps_popularity}
\end{table*}

\section{Evaluation and Results}
\label{sec:eval}
In this section, we present our CoD-based mobile app review summarization approach and we describe our evaluation procedure, including the experimental baselines, evaluation metrics, and key findings.

\subsection{Generating Review Summaries using the CoD Prompt}
The original CoD prompt is engineered to summarize news articles which are often written in the form of a story with a well-defined flow. However, app reviews are syntactically and semantically restricted in comparison to news articles. They also encompass the views of a large number of app users (authors). To workaround these restrictions, we adjust the CoD prompt as follows:

\begin{itemize}
	\item  To establish context, we direct the model to assume the role of an experienced review summarizer right from the beginning. Formally, we added the opening \textit{``You are an experienced summarizer of mobile app reviews,''} changed the task to \textit{``You will generate increasingly concise, entity-dense summaries of the above REVIEWS,''} and changed instructions such as \textit{``relevant to the main story''} to \textit{``relevant to the app functionality.''}
	
	\item We added a definition of what we consider an entity in mobile app user reviews. In particular, we defined an entity as \textit{``any functional or non-functional feature of the app that users mention in their reviews and perceive to either harm or enhance their overall experience.''} This definition is crafted to point the prompt toward app features, thus avoiding irrelevant entities that may appear in reviews. To prevent an over-representation of the negative or positive entities, we use \textit{``harm or enhance  user experience.''}   
	
	\item Reviews may contain irrelevant information such as people's names and URLs. They may also contain other apps' names, which can be confusing if appeared in an app's summary. To avoid fusing these entities in generated summaries, we added the clause, \textit{``Avoid apps' names other than \{\{app\}\} in summaries. Avoid specific information like locations, numbers, URLs, and emails in summaries.''}
	
	\item The CoD prompt restricts summaries' length to 80 words. While this limitation may suffice for news articles containing a uniform perspective of a single author, user reviews present a diverse range of viewpoints from a multitude of users (authors). The optimal length for an online review and its relation to helpfulness is heavily studied in the literature. For instance, in their study of online reviews, Huang et al.~\cite{Huang15} reported that when the review length exceeded 144 words, the relationship between word count and helpfulness became statistically insignificant. In our dataset, the average review length is around $105\pm15$ words, thus we restrict our summaries to 120 words. This length can help the summary capture the essence of users' diverse perspectives without being too brief or overly detailed. In other words, generated summaries can strike a balance between being comprehensive and cognitively digestible while maintaining an average length that users expect in the context of mobile app reviews~\cite{Koh22}.	
	\end{itemize}

\begin{figure}[ht]
	\scriptsize
	\begin{tcolorbox}[colback=gray!5,colframe=gray!40!black,]
		
		\texttt{Review: REVIEWS}\\
		
		\texttt{You will generate increasingly concise, entity-dense summaries of the reviews of the \{\{app\}\} app.}\\
		
		\texttt{\textbf{Repeat the following 2 steps 5 times.}}
		
		\begin{enumerate}
			\item \texttt{Identify 1-3 informative entities (";" delimited) from the reviews which are missing from the previously generated summary.}
			\item \texttt{Write a new, denser summary of identical length which covers every entity and detail from the previous summary plus the missing entities.} \\
		\end{enumerate} 
		
		\texttt{Note: An entity is any functional or non-functional feature of the app that users mention in their reviews and perceive to either harm or enhance their overall experience.} \\

		\textbf{\texttt{A missing entity is:}}
		
		\begin{itemize}
			\item \texttt{relevant to the app's functionality,}
			\item \texttt{specific yet concise (5 words or fewer),}
			\item \texttt{novel (not in the previous summary),}
			\item \texttt{faithful (present in the reviews),}
			\item \texttt{anywhere (can be located anywhere in the reviews).} \\
		\end{itemize}
		
		\textbf{\texttt{Guidelines:}}
		
		\texttt{The first summary should be long (4-5 sentences, ~120 words) yet highly non-specific, containing little information beyond the entities marked as missing. Use overly verbose language and fillers (e.g., "the review includes") to reach ~120 words. Make every word count: rewrite the previous summary to improve flow and make space for additional entities. Make space with fusion, compression, and removal of uninformative phrases like, "Users discuss." The summaries should become highly dense and concise yet self-contained, i.e., easily understood without the reviews. Missing entities can appear anywhere in the new summary. Never drop entities from the previous summary. If space cannot be made, add fewer new entities. Remember, use the same number of words for each summary. Avoid apps' names other than \{\{app\}\} in summaries. Avoid including personal and location specific information, like name, place, URLs, and emails, in summaries.}
	\end{tcolorbox}
	\caption{The CoD\textsubscript{r} prompt we used for app review summarization.}
	\label{fig:codr_prompt}
\end{figure}

We refer to our modified prompt as the CoD\textsubscript{r} prompt. The full text of the prompt is shown in Fig.~\ref{fig:codr_prompt}. Following Adams et al.~\cite{Adams23}, we use GPT-4 to generate our summaries. To manage computational resources, OpenAI imposes a restriction of 128k tokens on the size of text that can be processed by GPT in a single API request. As a result, it is not possible to submit all app reviews in a single summarization command. To workaround this limitation, we have to identify the portion of reviews to be included in the summary, thus staying within the request size requirements of GPT. This \textit{extract-then-abstract} approach is commonly used in text summarization to identify content that provides the most semantic value to the summarization process~\cite{Pilault20,Dou21}. 

To extract informative reviews, we first filter out non-English reviews. We use the \texttt{spacy-langdetect} package to detect the language of each review~\cite{Vasiliev20}. Reviews that are classified as English (en) are selected. The number of English reviews for each app is shown in Table~\ref{Tab:reviews_dataset}. We then implement a standard review ranking procedure to identify the most informative reviews for each app. Specifically, we remove noise terms such as HTML tags, emojis, punctuations, digits, and URLs from review text using regular expressions. We then remove English stop words (e.g., \textit{the}, \textit{a}, \textit{are}) based on the Natural Langauge Toolkit's (NLTK) list of stop words~\cite{Loper02}. The remaining words are then lower-cased and converted to their root forms using the WordNet lemmatizer. This procedure reduces reviews to vectors of unique words (bag-of-words). 

To quantify the information value of each review, we use Hybrid TF.IDF. This technique is commonly used in text processing to identify informative parts of short online text, such as user reviews or tweets~\cite{Ebrahimi23,Williams17}. The TF.IDF score of a word in a review is calculated as the product of the frequency of the word in the review (TF) and the word's scarcity across all reviews (IDF). Formally: 

\begin{equation}\label{Eq:tfidf_formula}
	\text{TF.IDF}(t, d, D) = \text{f}(t, d) \times \log\frac{N}{\text{df}(t, D)}
\end{equation}

In Eq.~\ref{Eq:tfidf_formula}, $t$ is a word, $d$ is a review which contains $t$, $D$ is a review collection of size $N$. The term $\text{f}(t, d)$ is the frequency of term $t$ in the review $d$ and $df(t,D)$ is the number of reviews in $D$ that contains the word $t$. The importance of a review is then calculated as the average Hybrid TF.IDF of its constituent words. 

Once reviews are ranked based on their TF.IDF scores, statistically representative samples from these reviews are generated. It is important for such samples to capture the diverse opinions of users as they appear in the reviews~\cite{Martin15}. In particular, user reviews are accompanied by 1-5 star ratings, reflecting the overall user impression of the app. Lower rating reviews are usually associated with negative sentiment, including complaints about app crashes and poor user experience, while reviews with higher star ratings usually feature praise for the app or descriptions of features that users find to enhance their experience~\cite{Maalej15,Tushev22}. For a summary to be unbiased towards either positive or negative reviews, it has to encompass the various viewpoints in the ground-truth~\cite{Maalej15,Khalid15}.  

Given that app reviews are non-uniformly distributed across the ratings, we use stratified sampling to generate a balanced sample of reviews for each app in our dataset~\cite{Maalej15}. The five star rating levels are considered as strata, where each rating is a stratum (single subcategory). We set the sample size to 350 reviews to maintain a 95\% confidence level at 5\% margin of error~\cite{Israel2009}. Using 350 as the total sample size (K), we compute the stratum size of reviews for each rating such that it reflects the corresponding stratum's size in the population of app reviews. The highest TF.IDF ranked reviews in each group are then selected to be included in the sample. Table~\ref{Tab:reviews_dataset} describes our dataset and the sample reviews distribution across the different star ratings for each of our studied apps. 

\begin{table}[ht]
	\centering
	\footnotesize
	\renewcommand{\arraystretch}{1.2}
	\caption{The star-rating distribution of the stratified review samples for the apps used in our analysis.}
	\smallskip 
	\begin{tabular}{p{0.12\textwidth} c c c c c c c}
		\Xhline{1.5\arrayrulewidth}
		\Xhline{1.5\arrayrulewidth}
		\multirow{2}{*}{\textbf{App}} & \multirow{2}{*}{\textbf{Reviews~(en)}} & \multicolumn{6}{c}{\textbf{Stratified Samples}} \\
		& & \textbf{\textcolor{black}{\large{$\filledstar$}}1} & \textbf{\textcolor{black}{\large{$\filledstar$}}2} & \textbf{\textcolor{black}{\large{$\filledstar$}}3} & \textbf{\textcolor{black}{\large{$\filledstar$}}4} & \textbf{\textcolor{black}{\large{$\filledstar$}}5} & \textbf{Total} \\
		\Xhline{1.5\arrayrulewidth}
		\Xhline{1.5\arrayrulewidth}
		Uber & 3,998 & 111 & 33 & 31 & 19 & 158 & 352 \\
		\rowcolor{gray!10} Lyft & 4,329 & 157 & 34 & 30 & 16 & 115 & 352 \\
		Tinder & 3,267 & 259 & 38 & 20 & 9 & 26  & 352 \\ 
		\rowcolor{gray!10} Bumble & 2,350 & 175 & 61 & 49 & 23 & 45 & 353 \\
		Robinhood & 5,009 & 162 & 24 & 20 & 31 & 116 & 353 \\
		\rowcolor{gray!10} Acorn & 1,454 & 224 & 32 & 20 & 14 & 63 & 353 \\
		Calm & 2,656 & 42 & 21 & 29 & 37 & 224 & 353 \\
		\rowcolor{gray!10} Headspace & 1,872 & 98 & 27 & 26 & 26 & 174 & 351 \\
		\Xhline{1.5\arrayrulewidth}
		\Xhline{1.5\arrayrulewidth}
		\rowcolor{gray!30} \textbf{Total} & \textbf{24,935} & \textbf{1,228} & \textbf{270} & \textbf{225} & \textbf{175} & \textbf{921} & \textbf{2,819} \\
		\Xhline{1.5\arrayrulewidth}
		\Xhline{1.5\arrayrulewidth}
	\end{tabular}
	\label{Tab:reviews_dataset}
\end{table}

We used OpenAI's $gpt-4-1106-preview$ model, the latest preview of the GPT-4 Turbo model at the time of conducting the study, to generate our summaries. To achieve consistent yet creative output, we set the $temperature$ parameter to 0.5. Lower $temperature$ values produce more focused and deterministic responses. Additionally, we moderated output variability using the top\_p parameter, which limits token selection to the smallest possible set whose cumulative probability exceeds a specified threshold. We set the top\_p value to 0.5 to limit the inclusion of potentially irrelevant or off-topic tokens while preserving a reasonable level of creativity. To further enhance output quality, we applied frequency and presence penalties of 0.1 to reduce token repetition in the generated summaries. As per OpenAI's documentation, the reasonable range for these penalties is between 0.1 and 1, where higher values strongly suppress repetition. We set the penalties closer to the lower end of this range.

\begin{figure*}[ht]
	\begin{subfigure}[b]{0.49\textwidth}
		\includegraphics[width=1\textwidth, trim={0 240 220 0}, clip, scale=1.1]{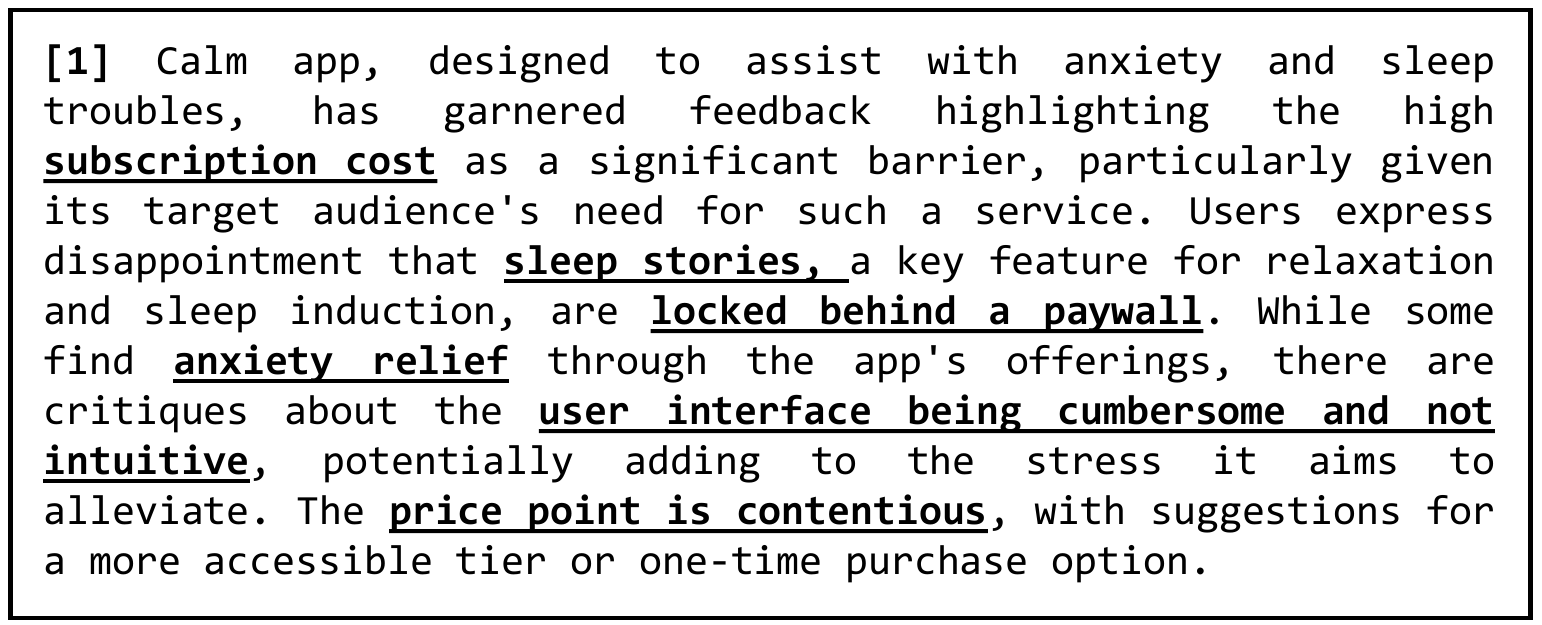}
		\label{fig:cod_itr_1}
	\end{subfigure}
	\hfill
	\begin{subfigure}[b]{0.49\textwidth}
		\includegraphics[width=1\textwidth, trim={0 240 220 0}, clip, scale=1.1]{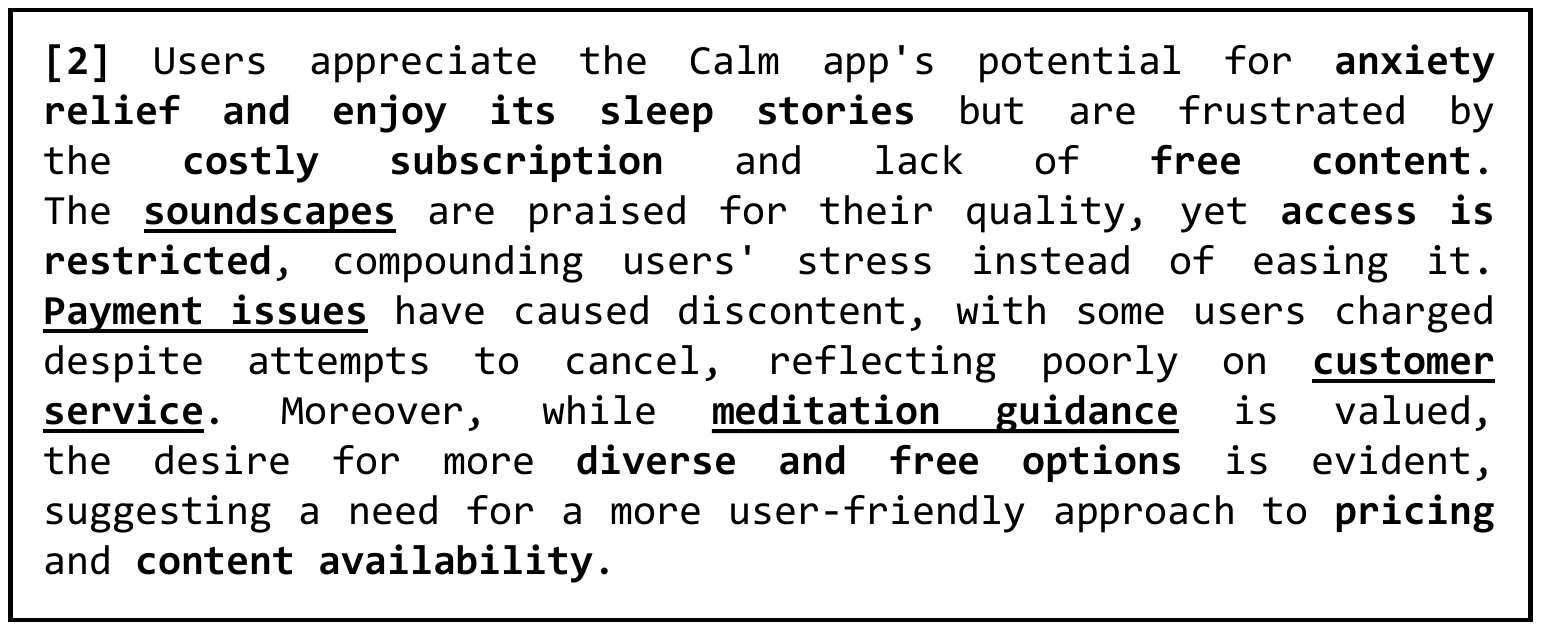}
		\label{fig:cod_itr_2}
	\end{subfigure}
	
	\begin{subfigure}[b]{0.49\textwidth}
		\includegraphics[width=1\textwidth, trim={0 270 220 0}, clip, scale=1.1]{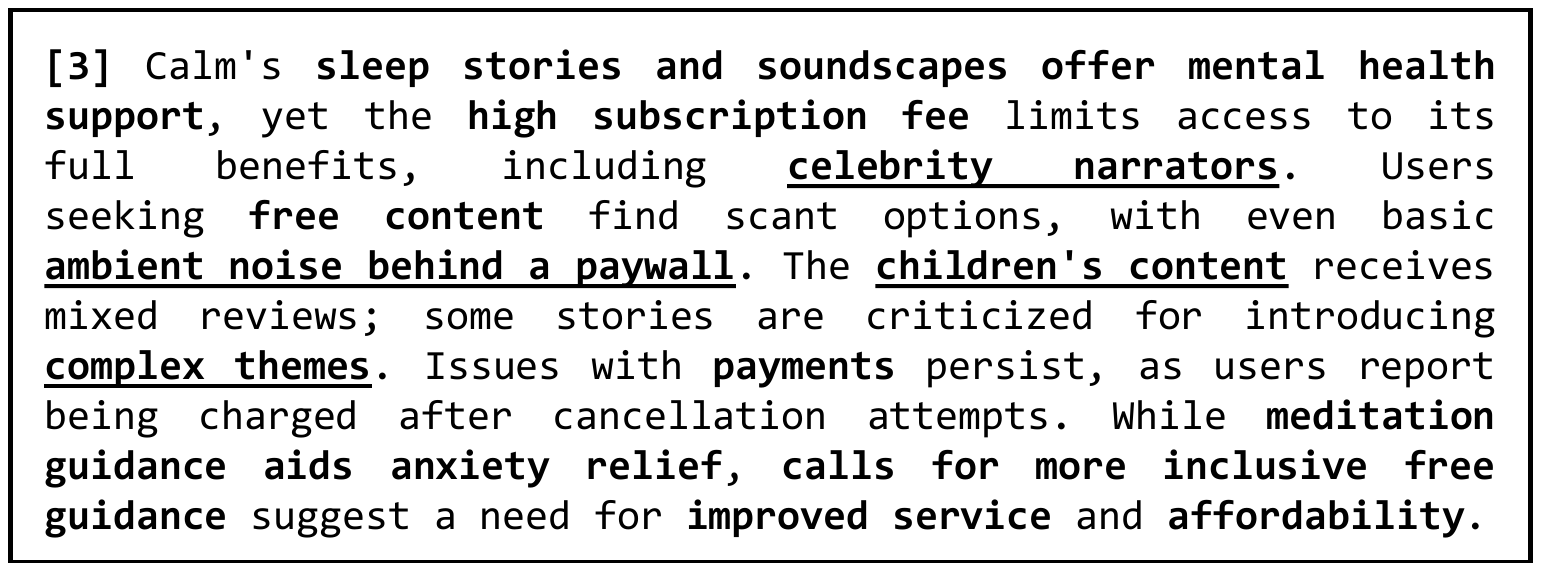}
		\label{fig:cod_itr_3}
	\end{subfigure}
	\hfill
	\begin{subfigure}[b]{0.49\textwidth}
		\includegraphics[width=1\textwidth, trim={0 270 220 0}, clip, scale=1.1]{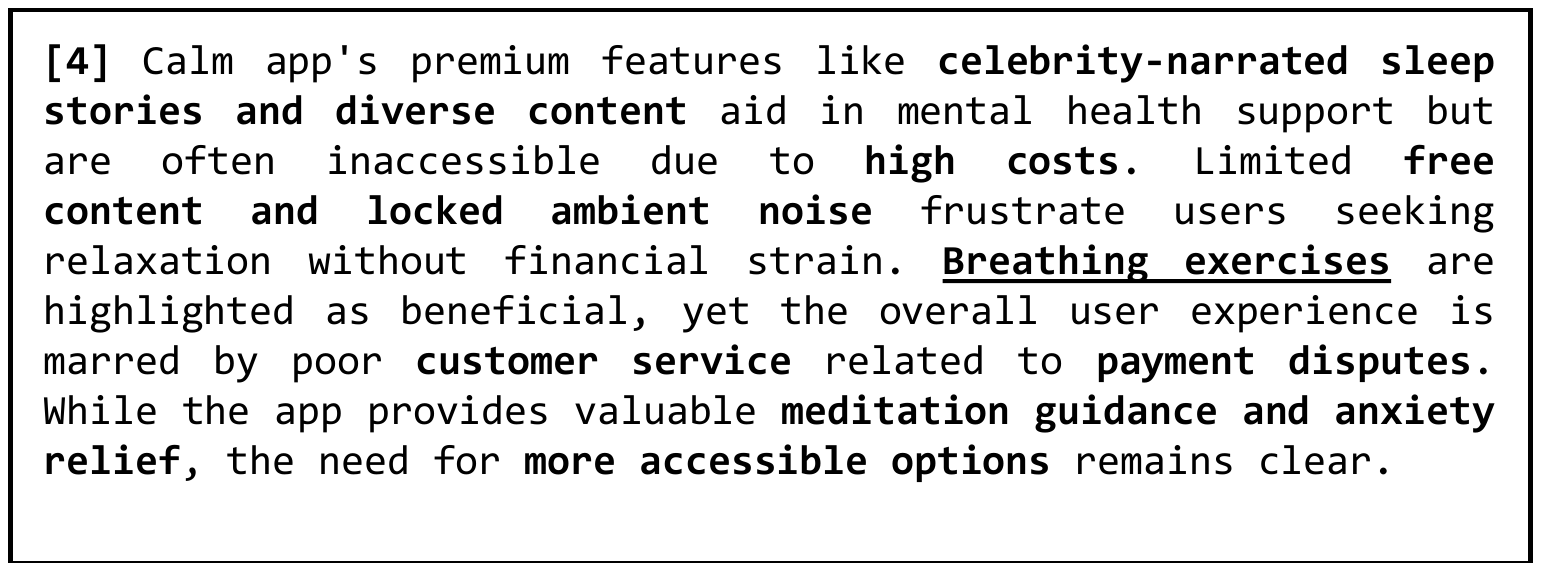}
		\label{fig:cod_itr_4}
	\end{subfigure}
	\caption{The review summaries generated for the mental health app Calm at the first four iterations of the CoD\textsubscript{r} prompt. Entities in the text are highlighted using bold text. Newly fused entities at each iteration are also underlined.}
	\label{Fig:iterative_summaries}
\end{figure*}

\textbf{Illustrative Example:} To illustrate the operation of  the CoD\textsubscript{r} prompt, we generated a summary of the reviews of the mental health app Calm. Fig.~\ref{Fig:iterative_summaries} shows the summaries generated at the first four iterations of the CoD\textsubscript{r} prompt. Entities in the summaries are shown in bold text while newly fused entities at each iteration are underlined. 
The example shows that the density of the summaries increases after each iteration. It also shows how the prompt paraphrases the text while fusing new entities. For example, the entity related to payments in the 3rd iteration summary, \textit{``Issues with payments persist, as users report being charged after cancellation attempts,''} is reduced to \textit{``payment disputes''} in higher iteration summaries to make space for new entities. 

\subsection{Experimental Baselines}
To assess the quality of the CoD\textsubscript{r} prompt's summaries, we experimented with two baselines, vanilla prompting and the extraction summarization technique Hybrid TF.IDF. These baselines can be described as follows:  

\begin{itemize}
	\item \textbf{Vanilla prompting:} Large Language Models (LLMs) have emerged as the state-of-the-art approach for text summarization, surpassing traditional extractive methods by generating coherent, context-aware summaries. They demonstrate superior performance in condensing information while maintaining factual accuracy and linguistic fluency~\cite{Jin24}. Vanilla prompting refers to a strategy where the LLM prompt is formulated to be simple and straightforward. The vanilla prompt can be used as a baseline to assess more sophisticated prompts. In our analysis, we use the vanilla prompt: \texttt{summarize the following app store reviews for the app <app\_name>. Make sure the summary does not exceed 120 words}. We enforce the 120-word length to avoid producing excessively long summaries~\cite{Liu23}.   
	
	\item \textbf{Hybrid TF.IDF summarization:} Hybrid TF.IDF is a frequency-based extractive summarization technique that is commonly used to summarize short and informal online text~\cite{Ebrahimi23,Elizabeth17}. Previous work showed that Hybrid TF.IDF constantly outperformed other extractive summarization techniques, such as SumBasic, Hybrid Term Frequency (TF), and LexRank in the context of mobile app user feedback~\cite{Jha18,Williams17,Elizabeth17}. Hybrid TF.IDF works initially by applying standard input text pre-processing to remove noise and reduce the chances of irrelevant text appearing in generated summaries. Individual sentences from each review are then extracted. This step is necessary to avoid always selecting abnormally large reviews that contain too many words. The Hybrid TF.IDF weights of individual words are then calculated using Eq.~\ref{Eq:tfidf_formula}. The importance of a review sentence is quantified as the average Hybrid TF.IDF weight of its constituent words. The \textit{k} sentences with the highest average Hybrid TF.IDF scores are then added to the summary. To minimize redundancy, or the chances of sentences describing similar issues appearing in the summary, the algorithm only adds sentences that share a similarity score that is less than a pre-defined threshold ($\lambda$) with other sentences that have already been added to the summary. A $\lambda < 0.1$ is commonly used in app review summarization~\cite{Ebrahimi23}. Sentences are added until the desired word length is met. To control for redundancy, we estimate the semantic similarity of reviews using their embedding vectors. In particular, we use GloVe ~\cite{Pennington14} to calculate the word embeddings of individual review sentences. Embeddings for word collections (e.g., sentences) can be computed by applying operations to their constituent word vectors, such as unweighted averaging or summation~\cite{Mikolov13-2}, as well as more advanced methods like Smooth Inverse Frequency (SIF)~\cite{Arora16}. In our analysis, we used the simple unweighted averaging method to obtain an embedding for each sentence in our review dataset~\cite{Arora16}. This method achieves effective approximations when the order of words in the text is unimportant.
\end{itemize}

\subsection{Evaluation Metrics}
Evaluating the performance of summarization techniques can be a challenging task~\cite{Fabbri21}. In general, summaries should be evaluated based on a combination of metrics that quantify their semantic density and readability (e.g., fluency, coherence, and informativeness). Summaries that accurately recall key entities from the source text and are easily comprehensible to the average target reader are considered higher in quality. To compare the performance of the CoD\textsubscript{r} prompt against the baselines, we employ two metrics, entity density and readability. In what follows, we describe these metrics in detail. 

\subsubsection{Entity Density}
\label{subsec:entity_density}
Entity density is a measure of semantic richness that is commonly used in text summarization tasks~\cite{Adams23}. In news article summarization, entities can be determined automatically, using tools such as the natural language processing library SpaCy~\cite{Vasiliev20}. The quality of a summary can be quantified as the ratio of entities to tokens in the summary's text. 
In the context of mobile app reviews, we define an entity as any functional or non-functional feature of the app that supports or harms its users experience. User experience is a product of their initial goals. In general, a user goal can be described as any abstract user objective that the system should achieve~\cite{Mylopoulos99}. For example, the goal of gig economy apps (e.g., Uber and Airbnb) is to foster social capital and economic growth in local communities~\cite{Rogers15,Fradkin15}, while the goal of fitness apps is to promote healthy lifestyles and sedentary behaviors in children and adults~\cite{Longyear21}. User goals do not have a clear-cut criterion for their satisfaction, however, they can be partially met. Given these assumptions, any aspect of the app that is perceived to support or harm these goals is a candidate to be included in the summary.  

We begin our evaluation by quantifying the semantic density of generated summaries. For each app, we generated 12 summaries: a vanilla prompt summary, Hybrid TF.IDF summary, five summaries generated by the CoD\textsubscript{r} prompt (one for each iteration), and five summaries generated by the original CoD prompt. We then calculate the semantic density of each summary using a systematic coding process. In particular, three independent judges examined each summary of each app individually, a total of 8 x 12 = 96 summaries. All three judges hold professional degrees in software engineering with an average of 4.2 years of professional experience in mobile app design and development. The task is to identify entities in the reviews. A pilot annotation session was held to help the annotators get a better sense of the entities available in the summaries. This session included annotating four summaries from two different apps. 

The manual labeling process concluded in three full weeks. Individual results were iteratively consolidated. Conflicts in the manual annotations were few and mainly related to the granularity of the entity. For example, in some of the Bumble app's summaries, safety concerns and fake profiles appeared as two separate entities. In other summaries, they appeared as one concern in statements such as \textit{``there are also safety concerns due to encounters with fake profiles.''} These conflicts were resolved after further discussions. The same process was followed for all other summaries with similar conflicts. The results of our manual annotation are provided in our supplemental material. 

\begin{figure*}[th]
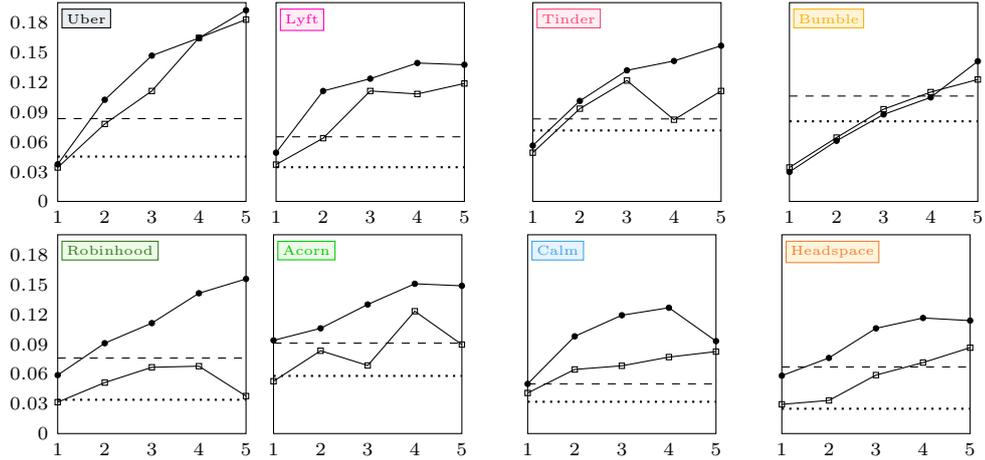

	\centering 
	\foreach \app\tfidf\vanilla\showlegend\forecolor\backcolor\bordercolor in \EntityDensityApps{
		\centering   
		\edef\temp{
			\noexpand\densityplot{data/entitydensity/nashcod/\app.csv}{\app}{data/entitydensity/adamscod/\app.csv}{(1, \vanilla) (5, \vanilla)}{(1, \tfidf) (5, \tfidf)}{\showlegend}{\forecolor}{\backcolor}{\bordercolor}
		}\temp	
	}
	
	\caption{The entity density (y-axis) of each summary generated for each individual app. The dashed line is the entity density of the vanilla prompt summary. The dotted line is the entity density of the summaries generated by Hybrid TF.IDF. The solid dot-marked line shows the density for each iteration (x-axis) of the CoD\textsubscript{r} prompt while the square-marked line shows the density for each iteration of the original CoD prompt.}
	\label{fig:entity_density_apps}
\end{figure*}

Table~\ref{Tab:entities_summ} shows the number of entities identified in each summary generated for each of our sample apps. We further plot the density of the different summaries for each app in Fig.~\ref{fig:entity_density_apps}. Density is calculated as the number of entities detected in the summary divided by the total number of words (tokens) in the summary~\cite{Adams23}. To enable a fair comparison, the CoD and the vanilla prompts are instructed to keep the length around 120 words, while for Hybrid TF.IDF review sentences are added to the summary until the length approaches 120 words. We executed a paired t-test to pairwise compare the density of summaries generated by different techniques (Table~\ref{tab:stats}). These results can be interpreted as follows: 

\begin{itemize}
	
	\item \textbf{CoD vs. CoD\textsubscript{r}:} Table~\ref{Tab:entities_summ} shows that the CoD\textsubscript{r} prompt was able to significantly outperform the original CoD prompt at the 3rd, 4th, and 5th iterations. At the 5th iteration, the CoD\textsubscript{r} prompt retrieved on average 2.63 more entities than the CoD prompt. The density diagrams also show that in the majority of apps, the CoD\textsubscript{r} prompt retrieved denser summaries for all iterations. In general, our results show that restricting entities to app features leads the summaries to exhibit more density. Consider, for example, the two following review summaries generated for the app Robinhood at the 3rd iteration of the CoD and CoD\textsubscript{r} prompts. 
	
	\begin{framed}
		\footnotesize
		$[$CoD\textsubscript{3}$]$ While Robinhood's user \textbf{interface is praised} for its simplicity, aiding in option trading for beginners, the app has been embroiled in controversy over \textbf{security concerns} and a class action lawsuit. The lawsuit stems from users' inability to buy certain stocks, perceived as market manipulation. Security issues include \textbf{data breaches} and questionable \textbf{handling of sensitive information}. These incidents have led to a loss of \textbf{trust} among some users, despite the platform's efforts to democratize stock trading.
	\end{framed}
	
	\begin{framed}
		\footnotesize
		$[$CoD\textsubscript{r3}$]$ Robinhood's \textbf{user interface} is lauded for ease of use, while its \textbf{educational resources} help demystify investing. However, users report \textbf{glitches and execution delays} that hinder timely trades, intensifying during market volatility and raising accusations of market manipulation. \textbf{Withdrawal} and \textbf{deposit} issues persist, with users feeling hamstrung by \textbf{restrictions} and \textbf{slow fund transfers}. \textbf{Customer service} is criticized for not resolving these urgent financial concerns effectively. \textbf{Options trading} is available but marred by the app's instability, especially during critical trading windows, detracting from the benefits of \textbf{instant deposits} and \textbf{crypto trading} features.
	\end{framed}
	
	In the above example, the CoD prompt picked up a security entity from the source reviews and elaborated on that entity using several sub-entities, including data breaches, handling of sensitive information, and trust. The summary went further to discuss a security lawsuit against Robinhood. This particular entity frequently appeared in the raw reviews of the app and seemed to substantially impact users' perception of Robinhood. However, this elaboration came at the expense of ignoring other important entities that the CoD\textsubscript{r} prompt was able to capture at the same iteration, including app glitches, instant deposits, crypto trading, withdrawals, and transaction delays. In general, the modified description of an entity has directed the prompt to prioritize breadth over depth. Nonetheless, the CoD\textsubscript{r} prompt still managed to pick up the security lawsuit in the 5th iteration after other important entities were included. The lawsuit appeared in the sentence \textit{``Robinhood's intuitive interface and educational content are overshadowed by lawsuits alleging market manipulation and data security concerns.''}
	
	\item \textbf{CoD\textsubscript{r} vs. the vanilla prompt:} Our results show that after the 3rd iteration, the CoD\textsubscript{r} prompt managed to outperform the vanilla prompt. At the 5th iteration, the CoD\textsubscript{r} prompt retrieved on average 2.25 more entities, significantly outperforming the vanilla prompt. However, the vanilla prompt was able to produce slightly denser summaries (0.38+) than the original CoD prompt. An interesting case was the online dating app Bumble where all prompts achieved comparable results. A deeper look into the summaries revealed that few dominant entities, such as fake profiles, subscription fees, customer service, and safety, stood out by appearing very frequently in the majority of this app's reviews. This relatively uniform distribution of entities over reviews helped the vanilla prompt generate dense summaries. This observation was further supported by the fact that even the extractive frequency-based approach hybrid TF.IDF was able to capture these issues in its extracted sentences.     
	
	\item \textbf{Hybrid TF.IDF:} This extractive technique was the least effective, capturing significantly less entities than any other technique in its generated summaries. Even with redundancy control, extractive techniques usually fail to establish deep connections between sentences, leading to contradicting statements within the same summary. For example, the two contradicting statements: \textit{``the [Uber] driver was fantastic''} and \textit{``drivers are cheaters and thieves''} appeared in Uber's TF.IDF summary. Furthermore, while sentence redundancy control helped to capture more entities, the summaries still exhibited significant redundancy, leading to incohesive and inconclusive summaries, especially in cases where the sentence itself was redundant, such as \textit{``when you get banned, they don't tell you why you got banned.''} 
	
\end{itemize}

\begin{table}
	\centering
	\footnotesize
	\renewcommand{\arraystretch}{1.2}
	\caption{The number of entities identified in each app's summaries. Twelve summaries are generated for each app: five CoD summaries, five CoD\textsubscript{r} summaries, one summary from the vanilla prompt, and one Hybrid TF.IDF summary.}
	\smallskip 
	\begin{tabular}{l c c c c c | c c c c c | c c}
		\Xhline{1.5\arrayrulewidth}
		\Xhline{1.5\arrayrulewidth}
		\multirow{2}{*}{\textbf{App}} & \multicolumn{5}{c}{\textbf{CoD}} & \multicolumn{5}{c}{\textbf{CoD\textsubscript{r}}} & \multirow{2}{*}{\textbf{Vanilla}} & \multirow{2}{*}{\textbf{TF.IDF}} \\
		& 1st & 2nd & 3rd & 4th & 5th & 1st & 2nd & 3rd & 4th & 5th & & \\
		\Xhline{1.5\arrayrulewidth}
		\Xhline{1.5\arrayrulewidth}
		Uber & 3 & 6 & 9 & 13 & 15 & 4	& 9	& 11 & 12 & 15 & 10 & 6 \\
		\rowcolor{gray!10} Lyft & 5 & 	7 & 	12 & 	12 & 14 & 	5 & 	10 & 	10 & 11 & 	11 & 10 & 	4 \\
		Tinder & 5 & 7 & 9 & 6 & 8 & 5 & 9 & 12 & 13 & 13 & 9 & 9 \\ 
		\rowcolor{gray!10} Bumble & 4 & 7 & 9 & 11 & 13 & 3 & 5 & 7 & 9 & 12 & 13 & 10 \\
		Robinhood & 3 & 4 & 5 & 4 & 2 & 6 & 8 & 10 & 12 & 14 & 9 & 4 \\ 
		\rowcolor{gray!10} Acorn & 5 & 6 & 5 & 8 & 6 & 9 & 9 & 10 & 11 & 11 & 9 & 6 \\
		Calm & 4 & 6 & 6 & 7 & 8 & 5 & 9 & 10 & 10 & 8 & 8 & 6 \\
		\rowcolor{gray!10} Headspace & 3 & 3 & 5 & 6 & 7 & 6 & 7 & 9 & 10 & 10 & 8 & 5 \\
		\Xhline{1.5\arrayrulewidth}
		\Xhline{1.5\arrayrulewidth}
		\rowcolor{gray!30} \textbf{Avg.} & \textbf{4.00}& \textbf{5.75} & \textbf{7.50} & \textbf{8.37} & \textbf{9.12} & \textbf{5.37}& \textbf{8.25} & \textbf{9.87} & \textbf{11.00} & \textbf{11.75} & \textbf{9.50} & \textbf{6.25} \\
		\Xhline{1.5\arrayrulewidth}
		\Xhline{1.5\arrayrulewidth}
		
	\end{tabular}
	\label{Tab:entities_summ}
\end{table}

\begin{table}[h]
	\centering
	\footnotesize
	\renewcommand{\arraystretch}{1.25}
	\caption{The results of the paired t-tests for the mean difference in entity density between summaries. Significance is measured at $p < 0.05$, \textbf{**} indicates a 99\% confidence and \textbf{*} indicates a 95\% confidence.}
	\smallskip
	\begin{tabular}{l c c c c c}
		\Xhline{1.5\arrayrulewidth}
		\Xhline{1.5\arrayrulewidth}
		\textbf{} & {CoD\_3} & {CoD\_4} & {CoD\_5} & {Vanilla} & {Hybrid TF.IDF} \\
		\Xhline{1.5\arrayrulewidth}
		\Xhline{1.5\arrayrulewidth}
		{CoDr\_3} & \textbf{0.006**} & 0.099 & 0.280 & \textbf{0.006**} & \textbf{0.000**} \\
		\rowcolor{gray!10} {CoDr\_4} & \textbf{0.001**} & \textbf{0.008**} & 0.061 & \textbf{0.000**} & \textbf{0.000**}\\
		{CoDr\_5} & \textbf{0.000**} & \textbf{0.002**} & \textbf{0.020*} & \textbf{0.002**} & \textbf{0.000**}\\
		\Xhline{1.5\arrayrulewidth}
		\Xhline{1.5\arrayrulewidth}
	\end{tabular}
	\label{tab:stats}
\end{table}

To get insights into our density analysis, we measure whether the CoD\textsubscript{r} prompt managed to pick up the dominant themes in the reviews, in other words, achieve high recall. To conduct such analysis, we sampled 30 reviews from the set of 300 reviews sampled for each app. Stratified sampling was used to generate representative samples. Three judges then independently examined each review in each sample, identifying the main entities in the reviews following the systematic procedure described earlier. Entities were iteratively identified and merged until saturation (i.e., no more entities were found). The recall of a generated summary was then calculated as the percentage of entities that appeared in the summary from the set of entities that were manually identified in the 30 review sample. 

\begin{figure} [ht]
	\centering
	\includegraphics[trim={2cm 2cm 0cm 1cm}, scale=.34]{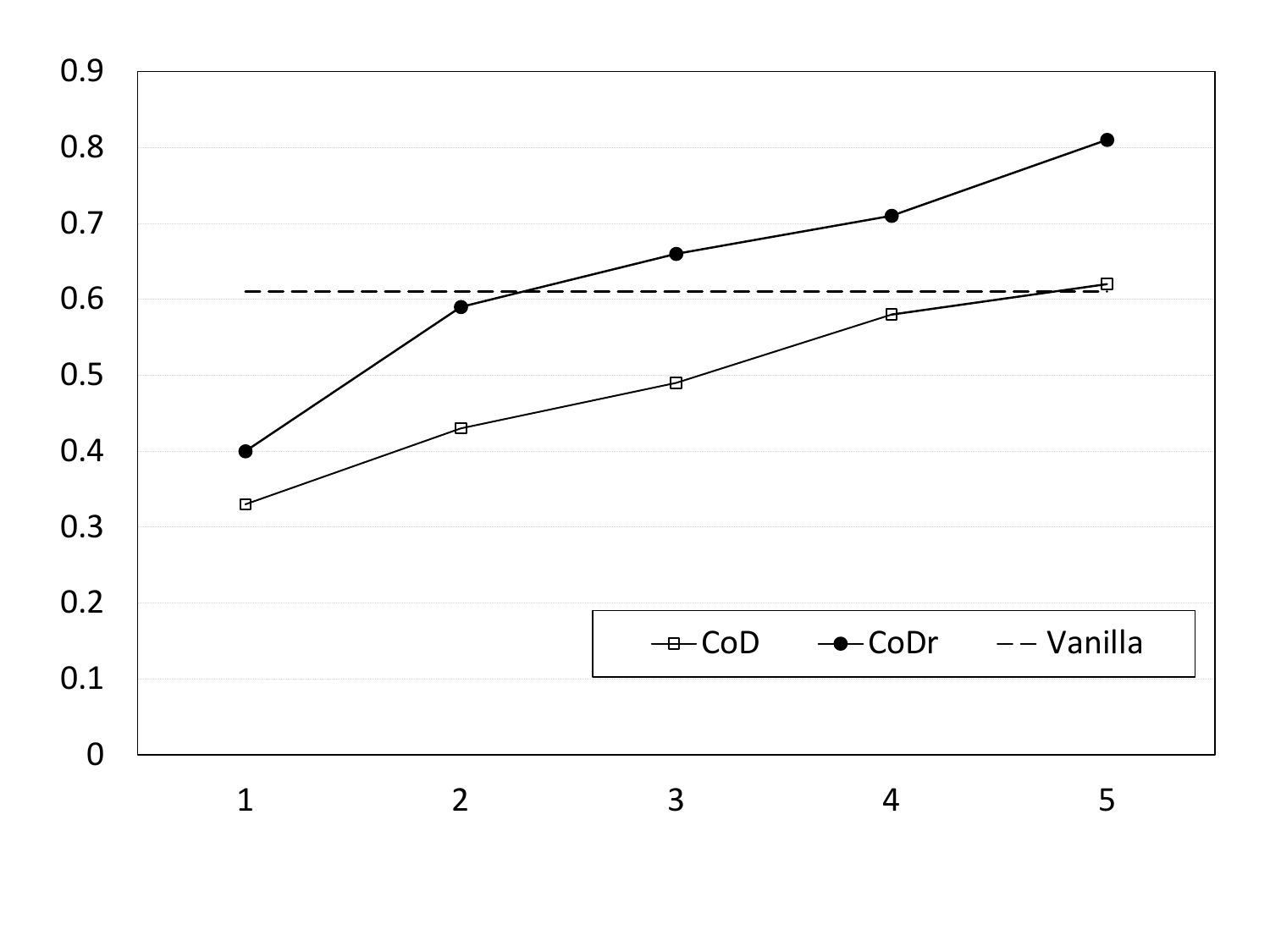}
	\centering
	\caption{Comparing the average recall (y-axis) of summaries generated at each iteration (x-axis) of the CoD and CoD\textsubscript{r} prompts as well as the summary generated by the vanilla prompt.}
	\label{Fig:recall}
\end{figure}

The results of the recall analysis are shown in Fig.~\ref{Fig:recall}. The figure shows that around the 3rd iteration, the CoD\textsubscript{r} prompt managed to capture more entities (81\%) than the vanilla prompt (64\%) while the original CoD prompt struggled to pick up more entities at higher iterations, settling at an average of 62\% recall at its 5th iteration. The difference in the number of recalled entities, even though small in some cases, can have a substantial influence on users' perspectives of the app. Specifically, given that extracted entities are representatives of prominent themes in the reviews, each entity could encompass the viewpoints of hundreds or even thousands of users. For instance, in the online dating app Tinder, only the CoD\textsubscript{r} prompt managed to capture concerns of transphobia, harassment, and sexual content solicitation in its summary. A search for these entities in the set of Tinder reviews revealed that around 5\% of the summarized reviews reported users experiencing transphobia, harassment, and sexual content solicitation while using the app.

	\begin{tcolorbox}[colback=gray!5,colframe=gray!40!black,]
\textbf{RQ$_1$}: 
A modified version of the Chain of Density (CoD) prompt that includes a definition of an entity tailored to the context of mobile app reviews significantly improves the semantic density and entity recall of review summaries, outperforming the original CoD prompt, a vanilla prompt, and Hybrid TF-IDF.				
\end{tcolorbox}

\subsubsection{Readability}
For an abstractive summarization technique to be deemed effective, it must produce readable summaries. While density is essential for achieving recall, generating overly dense summaries may lead to compressed and challenging-to-interpret text, especially when a length restriction is imposed. 

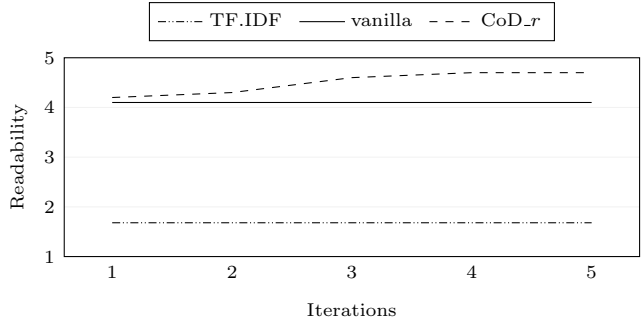
\begin{figure}
		\centering
		\begin{tikzpicture}
			\begin{axis}[
				title={},
				xlabel={Iterations},
				title style={font=\footnotesize},
				ylabel={Readability},
				label style={font=\footnotesize},
				ticklabel style={font=\footnotesize},
				grid=major,
				ymajorgrids=true,
				xmajorgrids=false,
				grid style={solid, gray!10},
				width=.7\textwidth, 
				height=120pt,
				y tick label style={/pgf/number format/fixed},
				ylabel style = {
					yshift = -18pt,
				},
				xlabel style = {
					yshift = 1pt,
				},
				ytick distance={1},
				ymin=1,
				ymax=5,
				tick style={
					draw=none,
				},
				legend style={at={(0.5, 1.28)}, 
					font=\footnotesize, 
					anchor=north,
					/tikz/every even column/.append style={column sep=0.15cm}, 
					align=center, 
					legend columns=-1}, 
				]
				
				\addplot[densely dashdotdotted]
				table [x=itr, y=tfidf, col sep=comma] {data/gpt-coherence.csv};
				\addlegendentry{TF.IDF}
				
				\addplot[solid]
				table [x=itr, y=vanilla, col sep=comma] {data/gpt-coherence.csv};
				\addlegendentry{vanilla}
				
				\addplot[dashed]
				table [x=itr, y=cod, col sep=comma] {data/gpt-coherence.csv};
				
				\addlegendentry{CoD$\_r$}
				
			\end{axis}
		\end{tikzpicture}
	\caption{The average readabiliy of Hybrid TF.IDF, the vanilla prompt, and CoD\textsubscript{r} prompt summaries as estimated by the GPT-4o model.}
	\label{fig:gpt_readability}
\end{figure}

To get a preliminary sense of the readability of the generated summaries, we start by comparing the readability of the CoD\textsubscript{r} prompt summaries to the experimental baselines. Recent evidence has shown that LLM-based text evaluation measures can reliably assess the readability of automatically generated text, producing readability scores that exhibit relatively high correlation with human judgments and outperforming traditional readability metrics, such as BLEU and ROUGE on multiple large-scale benchmarks~\cite{Liu232,Fu23}. Using this approach, the readability of a summary can be assessed by prompting an LLM to assign the text summary a score on a 1-5 scale.  A prompt then directs the model to analyze the summary and assign it a rating (analyze-then-rate). The LLM is instructed to read the summary, identify awkward or unclear sentences, look for errors in subject-verb agreement, ensure sentences are well-formed and not run-ons or fragments, and evaluate word choices and semantic cohesion. Following~\cite{Liu232}, we use GPT-4o to run our preliminary readability analysis. Fig.~\ref{fig:gpt_readability} shows that the CoD\textsubscript{r} prompt summaries exhibited the most readability. The vanilla prompt summaries exhibited comparative readability. However, the prompt was unable to match the CoD\textsubscript{r} prompt at higher iterations where cohesion was maximized. 

While the automated readability evaluation provided initial evidence on the quality of CoD\textsubscript{r} summaries, human evaluation remains essential to evaluate the readability of summaries~\cite{Adams23,Ouyang22}. Since our summaries are intended for end-user consumption, readability should not deteriorate as summaries become more dense. To conduct our human evaluation, we recruited 48 participants. Our objective was to determine if the readability of the generated summary declined with the increase in density. Ideally, summaries should maintain adequate readability since the CoD\textsubscript{r} prompt is instructed to preserve the naturalness of the text as much as possible~\cite{Meyer03,Reed16}.       

Convenience sampling was used to recruit our study participants from the population of students at a college campus. College-age adults are among the most active users of mobile apps~\cite{Rosen17,Kumcagiz16}. They use apps from a broad range of application domains, including social media, health and fitness, gig economy, entertainment, and education. The goal of our evaluation is to assess and compare the readability of the summaries generated by the CoD\textsubscript{r} prompt as their density increases. To conduct our evaluation, each participant was presented with three CoD\textsubscript{r} summaries and asked to assign them a score based on their readability. A 4-point Likert scale was used: unreadable, somewhat readable, readable, and easy to read. Such a scale is commonly used in NLP research as a standard method for assessing the readability of text~\cite{Miniukovich19}. Only the 3rd, 4th, and 5th iteration summaries were included in the study to control extraneous variables such as boredom or fatigue issues that may arise from reading five dense summaries.  	

To counteract order effects, we randomize the order in which the summaries were presented to different study participants, where for each app, each summary appeared at least once as the first summary (e.g., \{3, 4, 5\}, \{4, 5, 3\}, \{5, 4, 3\}, \{3, 5, 4\}, \{4, 3, 5\}, and \{5, 3, 4\}). The study participants were initially debriefed about the purpose of the study. Each participant was then handed a sheet including the summaries of one app. The study was conducted over two sessions: the first included 32 participants, and the second included an additional 16 participants. No time constraint was enforced; however, the majority of our participants completed the study in less than 15 minutes. 

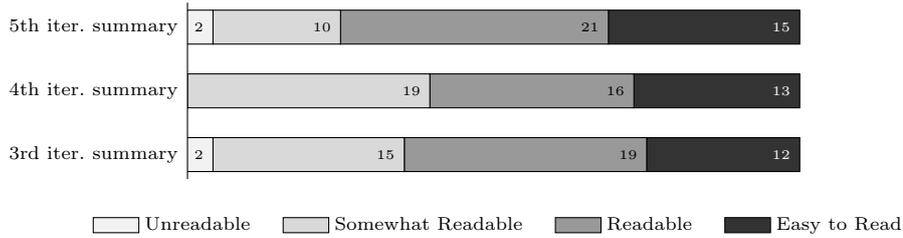
\begin{figure}
		\centering
		\begin{tikzpicture}
			\begin{axis}[
				xbar stacked,
				legend entries={\footnotesize{Unreadable}, \footnotesize{Somewhat Readable}, \footnotesize{Readable}, \footnotesize{Easy to Read}},
				 legend style={
					legend columns=-1,
					at={(xticklabel cs:0.5)},
					anchor=north,
					yshift=-10pt,
					draw=none,
					/tikz/every even column/.append style={column sep=10pt}
				},
				ytick=data,
				axis y line*=none,
				axis x line=none, 
				xtick style={draw=none}, 
				xtick=\empty,
				xticklabels=\empty,
				tick label style={font=\footnotesize},
				label style={font=\footnotesize},
				width=.75\textwidth,
				yticklabels={3rd iter. summary, 4th iter. summary, 5th iter. summary},
				xmin=0,
				xmax=49,
				area legend,
				y=6.5mm,
				bar width=4.5mm,
				enlarge y limits={abs=0.5},
				nodes near coords,
				nodes near coords align={horizontal},
				point meta=explicit symbolic,
				every node near coord/.append style={
					font=\footnotesize, 
					anchor=east, 
				},
				]
				\addplot[fill=gray!10] coordinates
				{(2,0) [\tiny{2}] (0,1.3) [] (2,2.6) [\tiny{2}]};
				\addplot[fill=gray!30] coordinates
				{(15,0) [\tiny{15}] (19,1.3) [\tiny{19}] (10,2.6) [\tiny{10}]};
				\addplot[fill=gray!80] coordinates
				{(19,0) [\tiny{19}] (16,1.3) [\tiny{16}] (21,2.6) [\tiny{21}]};
				\addplot[fill=black!80] coordinates
				{(12,0) [\textcolor{white}{\tiny{12}}] (13,1.3) [\textcolor{white}{\tiny{13}}] (15,2.6) [\textcolor{white}{\tiny{15}}]};
			\end{axis}
		\end{tikzpicture}
		\vspace{5pt} 
		\caption{The results of our readability study. The y-axis is the 3rd, 4th, and 5th iteration summaries of the CoD\textsubscript{r} prompt.}
		\label{Fig:participants_summary}
	\end{figure}

The results in Fig.~\ref{Fig:participants_summary} show that our participants found the summaries to be readable for the most part. In fact, contrary to our expectations, the 5th iteration summaries scored the most on the \textit{``easy to read''} scale. We use the Chi-Square test of independence to test for the statistical dependency between density and readability. Chi-square is a non-parametric statistic suitable for determining significant associations between categorical variables~\cite{Mchugh13}.

Our null hypothesis (H\textsubscript{0}) is that there is no relation between density and readability, or, increasing density does not significantly impact readability. Since we have three levels of density and four levels of readability, the degree of freedom of our test is set to 6 = (3-1) * (4-1) and the significance level is set to 0.05. The obtained Chi-square score is 5.80 with a \textit{p-value} $>$ 0.44, indicating that there is no sufficient evidence to reject the null hypothesis (H\textsubscript{0}). 

\begin{tcolorbox}[colback=gray!5,colframe=gray!40!black,]
\textbf{RQ$_2$}: There is no statistically significant evidence that readability declines when the density of generated summaries increases at higher iterations of the CoD\textsubscript{r} prompt. In fact, the relationship is pointing in the opposite direction, readability increases as density increases. 
\end{tcolorbox}

\subsection{Evaluating the CoD\textsubscript{r} Prompt on Other LLMs}
Given the rapid pace at which LLMs are evolving, it is important to assess the performance of the CoD\textsubscript{r} prompt on other emerging LLMs (\textbf{RQ$_3$}). This evaluation can help us gain further insights into the effectiveness of the prompt, quantify its ability to adapt to different LLMs, and inform prompt design for diverse applications and contexts.

To conduct our analysis, we experiment with two new LLMs, Gemini-1.5-Flash and Llama-3.1-70B-Instruct. The former was commercially released by Google in May 2024. The model supports 1M input tokens, suitable for providing larger context windows~\cite{Reid24}. The latter is an open source model that was released by Meta in July 2024. It contains 70B parameters and supports 128K context window length. These LLMs have been showing competitive performance compared to GPT-4 across several text generation tasks and over multiple benchmarks~\cite{Reid24,Ji24,Wang24}. 

\begin{figure*}[ht]
	\begin{subfigure}{0.5\textwidth}
		\centering
		\begin{tikzpicture}
    			\begin{axis}[
				title={},
        		xlabel={Iterations},
        		title style={font=\footnotesize},
        		ylabel={Avg. number of entities},
				label style={font=\footnotesize},
				ticklabel style={font=\footnotesize},
        		grid=major,
        		ymajorgrids=true,
        		xmajorgrids=false,
        		grid style={solid, gray!10},
        		width=1\textwidth, 
        		height=120pt,
				y tick label style={/pgf/number format/fixed},
				ylabel style = {
					yshift = -15pt,
				},
				xlabel style = {
					yshift = 1pt,
				},
				ytick distance={3},
				ymin=0,
				ymax=18,
				tick style={
					draw=none,
				},
				legend style={at={(0.4, 1.35)}, 
					font=\footnotesize, 
					anchor=north west,
					/tikz/every even column/.append style={column sep=0.2cm}, 
					align=left, 
					legend columns=-1},
  			 ]
       				 \addplot[solid]
       				 table [x=itr, y=gpt, col sep=comma] {data/all-llms-avg-ents.csv};
       				 \addlegendentry{GPT-4}
       				 
       				  \addplot[dashed]
       				 table [x=itr, y=gemini, col sep=comma] {data/all-llms-avg-ents.csv};
       				 \addlegendentry{Gemini-1.5-Flash}
       				 
       				  \addplot[densely dashdotdotted]
       				 table [x=itr, y=llama, col sep=comma] {data/all-llms-avg-ents.csv};
       				 \addlegendentry{Llama-3.1-70B-Instruct}
       				 
    			\end{axis}
		\end{tikzpicture}
	\end{subfigure}
	\begin{subfigure}{0.5\textwidth}
		\begin{tikzpicture}
    			\begin{axis}[
				title={},
        		title style={font=\footnotesize},
        		xlabel={Iterations},
        		ylabel={Avg. entity density},
				label style={font=\footnotesize},
				ticklabel style={font=\footnotesize},
        		grid=major,
        		ymajorgrids=true,
        		xmajorgrids=false,
        		grid style={solid, gray!10},
        		width=1\textwidth, 
        		height=120pt,
				ytick distance={0.045},
				ymin = 0,
				ymax = 0.28,
				y tick label style={/pgf/number format/fixed},
				ylabel style = {
					yshift = -7pt,
				},
				xlabel style = {
					yshift = 1pt,
				},
				tick style={
					draw=none,
				},
  			 ]
       				 \addplot[solid]
       				 table [x=itr, y=gpt, col sep=comma] {data/all-llms-avg-density.csv};
       				 
       				 \addplot[dashed]
       				 table [x=itr, y=gemini, col sep=comma] {data/all-llms-avg-density.csv};
       				 
       				 \addplot[densely dashdotdotted]
       				 table [x=itr, y=llama, col sep=comma] {data/all-llms-avg-density.csv};
       				 
    			\end{axis}
		\end{tikzpicture}
	\end{subfigure}
	\vspace{-8pt}
	\caption{Left is the average number of entities fused into the summaries and right is the average entity density of the summaries at the different iterations of the CoD\textsubscript{r} prompt across the different LLMs.}
	\label{fig:ents_density_avgs_oth_llms}
\end{figure*}
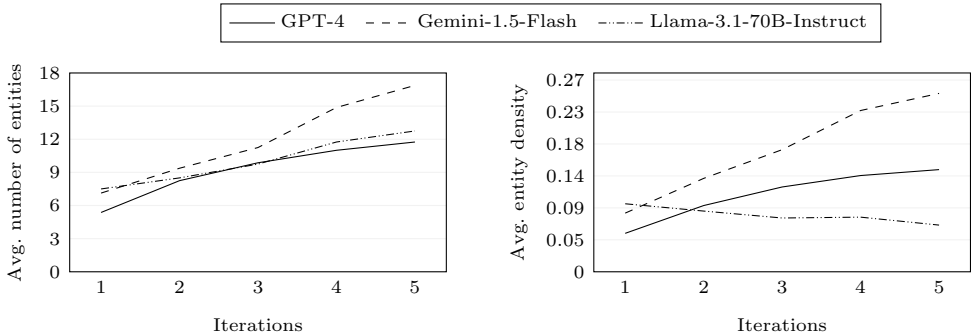

To conduct our analysis, we used the CoD\textsubscript{r} prompt (temperature = 0.5 and top\_p = 0.5) to prompt both LLMs to summarize the app reviews in our dataset. We then followed the same entity extraction procedure outlined in Section~\ref{subsec:entity_density} to quantify the density of generated summaries. Fig.~\ref{fig:ents_density_avgs_oth_llms} shows the average number of unique entities as well as the average entity density of summaries generated across the five iterations of the CoD\textsubscript{r} prompt for GPT-4, Gemini-1.5-Flash, and Llama-3.1-70B-Instruct. 

Our results show that the CoD\textsubscript{r} prompt successfully triggered all LLMs to iteratively increase the number of fused entities in generated summaries. We also observe that the number of extracted entities is substantially larger in Gemini-1.5-Flash and Llama-3.1-70B-Instruct summaries than GPT-4 summaries, especially at the higher iterations of the prompt. However, examining the density chart shows that the increase in the number of entities has drastically altered the structure of the summaries. 

Consider for example Fig.~\ref{fig:rb_summs_three_models} which shows the summaries generated at the 5th iteration of the CoD\textsubscript{r} prompt for the Headspace app using GPT-4, Gemini-1.5-Flash, and Llama-3.1-70B-Instruct. The Gemini-1.5-Flash summary is extremely dense with 19 entities. However, to comply with the length constraint, Gemini-1.5-Flash dropped words that conveyed sentiment and perception from the summary; describing entities as being \textit{noted}, \textit{discussed}, or \textit{mentioned} without any indication whether users perceived these entities positively or negatively. The density analysis also showed that the Llama-3.1-70B-Instruct summaries were packed with more entities than the GPT-4 summaries. However, they were excessively long. In other words, Llama-3.1-70B-Instruct relaxed the length constraint in order to maintain context and readability. In the example in Fig.~\ref{fig:rb_summs_three_models}, the Llama-3.1-70B-Instruct summary of Headspace reviews contained several words and phrases that conveyed perception, such as \textit{works best}, \textit{help users}, and \textit{excellent tool}.   	

In summary, our results show that the CoD\textsubscript{r} prompt managed to guide all LLMs to generate dense summaries that captured most entities in users reviews, however, generated summaries substantially varied in their structure for different LLMs. GPT-4 can strike a balance between length and density. It captures the most important entities in the reviews while maintaining readability and context to a large extent. If recall is the main concern, then Gemini-1.5-Flash can be used. However, generated summaries can be too dense to be informative as the model prioritizes the length constraint over readability. If the length of the summary is not an issue, then Llama-3.1-70B-Instruct can be an option. However, the generated summaries can be excessively long and mainly leaning towards the positive aspects of the app. Next, we discuss the implications of our findings.    

\begin{tcolorbox}[colback=gray!5,colframe=gray!40!black,]
\textbf{RQ$_3$}:  The CoD\textsubscript{r} prompt can guide LLMs such as Gemini-1.5-Flash and Llama-3.1-70B-Instruct to generate semantically dense summaries of mobile app reviews. However, generated summaries vary in their structure and length. GPT-4 can strike a balance between summaries' lengths and their semantic densities.
\end{tcolorbox}

\begin{figure}[ht]
	\begin{framed}
	\footnotesize
		$[$ \textbf{\texttt{Gemini-1.5-Flash}} $]$ The app offers \textbf{guided meditations, sleep aids, anger management}, and \textbf{child content}. \textbf{Privacy, in-app purchases, customer service,} and the \textbf{free trial} are discussed. \textbf{User interface issues, account deletion, accessibility problems, daily meditations, subscription costs, political content}, and \textbf{content variety} are noted. \textbf{Sleep sounds} and \textbf{meditation progress tracking} are mentioned. \textbf{App crashes} and the lack of \textbf{free content} are additional concerns.
	\end{framed}
	
	\begin{framed}
		\footnotesize
		$[$ \textbf{\texttt{GPT-4}} $]$ Headspace's \textbf{user interface design} and \textbf{device compatibility} issues create barriers to accessing a variety of high-quality \textbf{meditation content}. The app's focus on subscription revenue leads to \textbf{locked content} and \textbf{free trial} traps, causing {payment disputes} poorly handled by \textbf{customer service}. \textbf{In-app ads} and unnecessary distractions diminish the meditation experience. \textbf{Data privacy} concerns arise from aggressive user tracking, while \textbf{political content} alienates users seeking neutrality. Despite these issues, some users appreciate the app's potential for \textbf{diverse content} and its ability to aid in sleep and stress reduction.
	\end{framed}
	
	\begin{framed}
		\footnotesize
		$[$ \textbf{\texttt{Llama-3.1-70B-Instruct}} $]$  The app offers a wide range of meditation and mindfulness techniques, including \textbf{breathing exercises, sleep casts}, and \textbf{yoga}. Users can choose from various \textbf{meditation teachers} and \textbf{topics}, such as stress, sleep, and focus, to find what works best for them. The app also provides \textbf{personalized meditation plans} and \textbf{tracks progress} to help users stay consistent. With its \textbf{soothing atmosphere} and \textbf{gentle guidance}, this app is an excellent tool for anyone looking to reduce stress and improve their mental well-being. Additionally, the app offers sleep casts and yoga sessions to help users relax and unwind. The app also features \textbf{meditation courses, focus music}, and \textbf{mindful moments} to help users stay focused and present throughout the day. Furthermore, the app has a \textbf{community feature} that allows users to connect with others and share their meditation experiences. The app also offers \textbf{meditation sessions} specifically designed for kids, anxiety, sleep, stress, depression, PTSD, chronic pain, and addiction, making it a great resource for users of all ages and needs. With its comprehensive library of meditation sessions, this app is an excellent tool for anyone looking to improve their mental health and well-being.
	\end{framed}
	
	\caption{The summaries generated for the Headspace app at the 5th iteration of the CoD\textsubscript{r} prompt, using Gemini-1.5-Flash, GPT-4, and Llama-3.1-70B-Instruct.}
	\label{fig:rb_summs_three_models}	
\end{figure}

\section{Discussion}
\label{sec:discuss}
In this paper, we propose using LLMs to generate semantically-dense, natural, and human interpretable summaries of mobile app reviews. 
Our analysis revealed that the Chain of Density (CoD) prompt can effectively guide GPT-4 to generate abstractive summaries of mobile app reviews if provided with a clear definition of an \textit{entity} in app user feedback. The prompt (we referred to as CoD\textsubscript{r}) can correctly identify the salient themes in user reviews and consolidate them into a single summary that is intended for end-user consumption. The iterative fusing of new entities in the summary can be particularity effective in cases where entities are not uniformly distributed over reviews, overcoming the limitations of simpler baselines, such as extractive summarization or vanilla prompting (\textbf{RQ$_1$}). 

Our evaluation of the readability of generated review summaries revealed that the CoD\textsubscript{r} prompt managed to maintain decent levels of readability while increasing the semantic density of summaries. In fact, contrary to our expectations, the results of our human evaluation revealed that the readability slightly, but not significantly, improved for higher iterations (\textbf{RQ$_2$}). Our analysis also revealed that the density and structure of CoD\textsubscript{r} summaries vary across different LLMs. For example, when using Google's Gemini-1.5-Flash model, the CoD\textsubscript{r} prompt generated extremely dense summaries of mobile app reviews. However, generated summaries often lacked context, as words conveying emotions and perceptions were omitted to prioritize retrieved entities. Similarly, Meta's Llama-3.1-70B-Instruct model generated summaries with high density. However, the summaries were substantially longer as the model relaxed the length constraint to accommodate more entities. In contrast to these two models, OpenAI's GPT-4 struck a balance between summary density and length, enabling the CoD\textsubscript{r} prompt to generate more contextually rich and concise summaries (\textbf{RQ$_3$}).

Our work in this paper illustrates the effectiveness of LLM-based methods in the realm of unstructured and informal text of mobile app reviews. Through their underlying complex neural architecture, they can capture the diverse perspectives of a large number of users while mitigating expressiveness degradation~\cite{Jin24,Shrestha24}. In terms of latency and cost (API fees), we leveraged API services from OpenAI, Google, and Amazon Bedrock to access GPT-4, Gemini-1.5-Flash, and Llama-3.1-70B-Instruct models, respectively. Our results showed that, on average, GPT-4 and Gemini-1.5-Flash executed the CoD\textsubscript{r} prompt and generated summaries for each app in 2.2 and 2.9 seconds, respectively. Llama-3.1-70B-Instruct was slower, averaging 3.4 seconds per app. In conclusion, while all three models generate summaries within a few seconds, the differences in speed may impact real-time applications requiring faster responses. 

In term of API fees, OpenAI's GPT-4 charged \$2.5 per million input tokens and \$10 per million output tokens. In our setup, the API input included the task instructions (prompt) and 350 reviews per app, while the API output contained a JSON list of five summaries. Among the evaluated LLMs, GPT-4 incurred the highest cost, averaging \$0.102 per summary. In contrast, Gemini-1.5-Flash and Llama-3.1-70B-Instruct offered lower costs, staying below \$0.05 per summary. In general, while GPT-4 was the fastest, it was also the most expensive. Gemini-1.5-Flash and Llama-3.1-70B-Instruct offered cost-effective alternatives, though with slightly higher latency. These findings highlight a trade-off between latency, cost, and model selection, depending on application needs. 

\begin{figure}
	\centering
	\includegraphics[trim={0cm 0cm 0cm 0cm}, scale=.32]{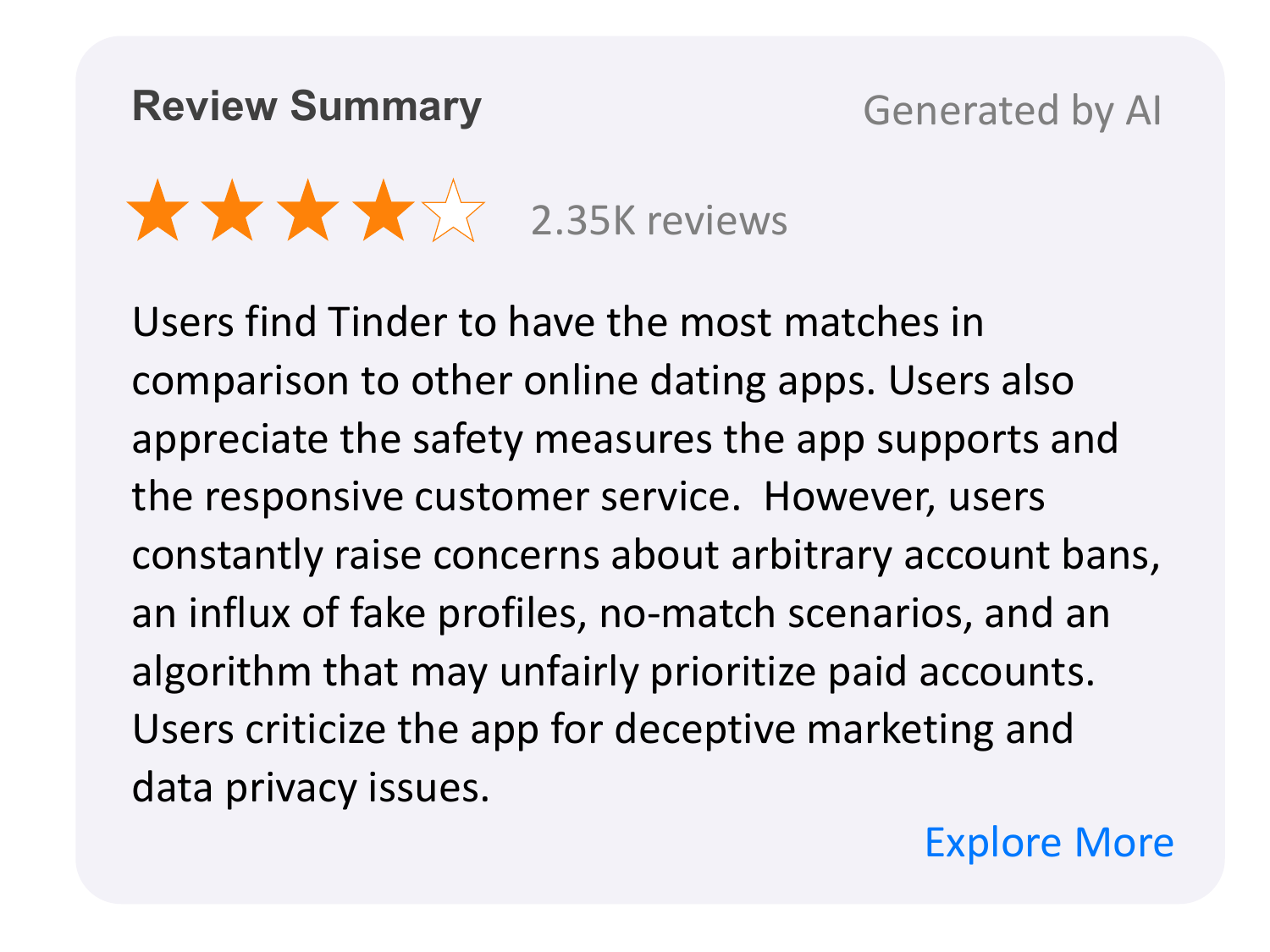}
	\centering
	\caption{A suggested interface for integrating review summaries into app store review feeds.}
	\label{Fig:integration}
\end{figure}

Fig.~\ref{Fig:integration} shows how the proposed summaries can be integrated into existing app store rating systems. Similar to online retail platforms, such as Amazon, a summary of user reviews along with the average rating can be displayed at the top of the review feed and users can opt to explore more. Our expectation is that these summaries can effectively inform app selection and installation decisions. In particular, users can learn about app features and their impact on their goals and experiences without reading through thousands of reviews. For instance, users of online dating apps can grasp how helpful an app is in finding compatible partners, fostering meaningful relationships, or preventing harassment, while users of investing apps can learn if an app can help them achieve their financial goals, generate personalized investing plans that meet their needs, or protect them from fraud. App developers can also detect shifts in their end-users' preferences, usage patterns, and pain points by tracking changes in generated summaries. Given this information, they can direct maintenance efforts to optimize their end-user experience, and ultimately, survive market selection. 

Integrating summaries into review feeds can effectively counter anchoring bias. Evidence from online retail shows that customers tend to prioritize the most recent product reviews, or those displayed at the top of the list, when making purchase decisions. This tendency is known as anchoring bias or effect. In behavioral economics, anchoring refers to the disproportionate influence of an initially presented value on decision-making~\cite{Tversky74}. App reviews in app stores are often displayed chronologically or ranked by the number of likes, causing users' perceptions to be influenced by the salience of top reviews. Displaying summaries that capture the opinions of a broad range of users at the top of the review feed can mitigate the anchoring effect of these prominent reviews~\cite{Hu19}. Moreover, presenting balanced summaries that highlight both positive and negative aspects of an app can help reduce negativity bias, or the tendency of consumers to weigh negative reviews more heavily than positive ones. Behavioral economics research indicates that negative reviews significantly impact consumer behavior because they are often perceived as more novel, attention-grabbing, and salient than positive reviews~\cite{Colmekcioglu22,Chevalier06}.

In terms of future enhancements, we anticipate that incorporating other types of review information into the prompt may help synthesize more accurate summaries. For instance, including factors such as the number of approvals (likes) a review has received may trigger the prompt to weigh entities in reviews differently. In addition, the date of the reviews and the credibility of the author (if available) can be experimented with. Such data points can be automatically extracted from app reviews, thus the overhead should be manageable. However, there is always a concern of producing overly complex prompts~\cite{Liu23}. 

\section{Threats to Validity}
\label{sec:threats}

The study presented in this paper has several limitations that could potentially jeopardize the validity of our results. One threat to the external validity of our study arises from the fact that we analyzed reviews from only eight apps. This may limit the generalizability of our findings to other apps. However, popular apps have larger user bases and often receive substantially more reviews, making summarization more necessary for these apps than for small apps that receive only a few reviews per day or week. In addition, we selected our apps from a diverse set of application domains to further enhance the external validity of our results.  

A threat to the internal validity of our study can originate from the  manual annotation of our review samples. To mitigate this threat, a pilot session was held at the beginning of the annotation process to address our annotators' concerns. Furthermore, to minimize fatigue, no time constraint was enforced. Overall, these measures helped preserve the integrity of the manually annotated data; a small conflict rate of $\sim5\%$ was detected between annotators. Several measures were also taken to control extraneous variables during our human evaluation study, including boredom, order effect, and mental fatigue.

Other internal validity threats might stem from the specific LLMs used in our analysis. Different models, different types of training (single or few-shots), or different values for the hyperparameters may lead to better results~\cite{Fan23}. Other threats may stem from the reliability and consistency of LLM responses~\cite{Sallou24}. However, given the level of sophistication of the CoD prompt (e.g., iterative nature, restrictions on length, and clear instructions), hallucinations are highly unlikely to occur.  A threat can also arise from the fact that we only experimented with a fixed-length summary of 120 words. We imposed such a constraint on the length to avoid generating excessively long summaries and minimize cognitive overload~\cite{Huang15}. Nonetheless, future independent studies should investigate the optimal length for app review summaries. 

\section{Conclusion and Future Work}
\label{sec:conc}
In this paper, we leveraged LLMs to generate abstractive summaries of mobile app reviews. In particular, we evaluated the effectiveness of the Chain-of-Density (CoD) prompt in summarizing representative samples of mobile app reviews collected from eight different popular apps. Our results showed that modifying the CoD prompt to recognize app features in user reviews helped generate comprehensive, increasingly dense, and interpretable summaries of reviews. The modified prompt, CoD\textsubscript{r}, was evaluated against multiple baselines. The results showed that it outperformed the original CoD prompt, a vanilla prompt, and the extractive summarization technique Hybrid TF.IDF in terms of semantic density and recall. A human evaluation with 48 study participants showed that the CoD\textsubscript{r} prompt maintained the readability of generated summaries while increasing their semantic density. 

We anticipate that the integration of natural summaries in the rating systems of modern mobile app stores will help app users make more informative app selection decisions as well as help app developers identify important entities of user feedback in their review streams. To evaluate these assumptions, in our future work, we will create mock-ups that resemble popular app stores and run user experiments in the form of randomized controlled trials. Large and heterogeneous samples of study participants will be recruited. Our study participants will be asked to install apps and recommend apps to their friends or peers. Through such studies, we will analyze the impact of providing review summaries on app user' information exploration and app selection decisions~\cite{Jasim22}, focusing on common consumer biases, such as the anchoring bias and negativity bias. 

\section*{Data Availability Statement}
Our data is publicly available\footnote{\url{https://figshare.com/s/e7a9127f7acb8bba9439}}. The data includes our set of reviews, our review samples, and the scripts we used to scrape, clean, and summarize the data. We also include the results of our readability evaluation study. 

\section*{Conflict of Interest}
The authors declare that they have no conflict of interest.

\bibliography{ase}

\end{document}